\documentclass[twocolumn,pra,aps,superscriptaddress]{revtex4}
\usepackage{subfigure}
\usepackage{psfrag,graphicx}
\usepackage{pict2e}
\usepackage{dcolumn}
\usepackage{amsmath,amssymb}
\usepackage{bm}
\usepackage{color}
\usepackage{latexsym}
\usepackage{epstopdf}
\usepackage{color}
\usepackage[english]{babel}

\usepackage{amsfonts}
\usepackage{bm}
\usepackage{natbib}
\usepackage{bbold}
\usepackage{enumerate}
\definecolor{myblue}{rgb}{0.2,0.2,0.8}
\definecolor{myzard}{cmyk}{0,0,0.05,0}
\definecolor{mywhite}{rgb}{1,1,1}
\definecolor{mywhite}{rgb}{1,1,1}
\definecolor{myred}{rgb}{1,0.,0.3}
\usepackage[colorlinks=true,citecolor=myblue,linkcolor=myred]{hyperref}

\def\one{\mathbb{1}}
\def\zero{\mathbb{0}}

\def\cc{{\rm c}}
\def\dd{{\rm d}}

\def\pp{{\rm p}}
\newcommand{\tr}[1]{\textcolor{black}{#1}}
\begin{document}
\title{Topological input-output theory for directional amplification}

\author{Tom\'as Ramos}
\affiliation{Instituto de F\'isica Fundamental, IFF-CSIC, Calle Serrano 113b, 28006 Madrid, Spain}
\affiliation{DAiTA Lab, Facultad de Estudios Interdisciplinarios, Universidad Mayor, Santiago, Chile}

\author{Juan Jos\'e Garc{\'i}a-Ripoll}
\affiliation{Instituto de F\'isica Fundamental, IFF-CSIC, Calle Serrano 113b, 28006 Madrid, Spain}

\author{Diego Porras}
\email{diego.porras@csic.es}
\affiliation{Instituto de F\'isica Fundamental, IFF-CSIC, Calle Serrano 113b, 28006 Madrid, Spain}

\date{\today}

\begin{abstract}
We present a topological approach to the input-output relations of photonic driven-dissipative lattices acting as directional amplifiers.
Our theory relies on a mapping from the optical non-Hermitian coupling matrix to an effective topological insulator Hamiltonian. 
This mapping is based on the singular value decomposition of non-Hermitian coupling matrices, whose inverse matrix determines the linear input-output response of the system. %
In topologically non-trivial regimes, the input-output response of the lattice is dominated by singular vectors with zero singular values that are the equivalent of zero-energy states in topological insulators, leading to directional amplification of a coherent input signal. 
In such {\it topological amplification regime}, our theoretical framework allows us to fully characterize the amplification properties of the quantum device such as gain, bandwidth, added noise, and noise-to-signal ratio. 
We exemplify our ideas in a one-dimensional non-reciprocal photonic lattice, for which we derive fully analytical predictions. We show that the directional amplification is near quantum-limited with a gain growing exponentially with system size $N$, while the noise-to-signal ratio is suppressed as $1/\sqrt{N}$. This points out to interesting applications of our theory for quantum signal amplification and single-photon detection.
\end{abstract}

\maketitle
\setcounter{tocdepth}{2}
\begingroup
\hypersetup{linkcolor=black}
\endgroup
\section{Introduction}
Topological photonics \cite{Lu2014,ozawa18rmp} is the research field that applies topology to the study of transport and control of light in systems such as photonic crystals, cavity arrays and metamaterials.
This field has been inspired by the physics of topological insulators, where electronic transport occurs through symmetry protected surface or edge states. 
The framework that describes those electronic materials is topological band theory (TBT), which allows for a classification of topologically non-trivial phases depending on dimensionality and symmetry \cite{Schnyder08,ryu10,Bansil16,Chiu16}.
Topological effects in photonic systems are not only interesting from a fundamental point of view, but they could also play a significant role in applications such as photon routing and amplification. 

After Haldane and Raghu's pioneering work \cite{Haldane08}, first realizations of topological phases were implemented in photonic spin Hall systems  \cite{Hafezi11}.
In the last decade photonic topological phases have been investigated by breaking time-reversal symmetry with magnetic fields  
\cite{Wang09,koch10,Anderson16,Lu2016,Owens18}
or periodic drivings \cite{fang12,Peropadre13,Rechtsman13,roushan17,sounas17,mukherjee18}. 
Analogous ideas have been explored in
optomechanical systems \cite{Hafezi12,Schmidt15,Ruesink16,Shen16,bernier18pra} or even in purely vibronic or mechanical systems \cite{Bermudez11prl,Bermudez12,Susstrunk15,Kiefer19prl},
as well as in spin-cavity setups \cite{harder17prb,zhang17natcomm}.

\begin{figure}[t]
	\includegraphics[width=0.4\textwidth]{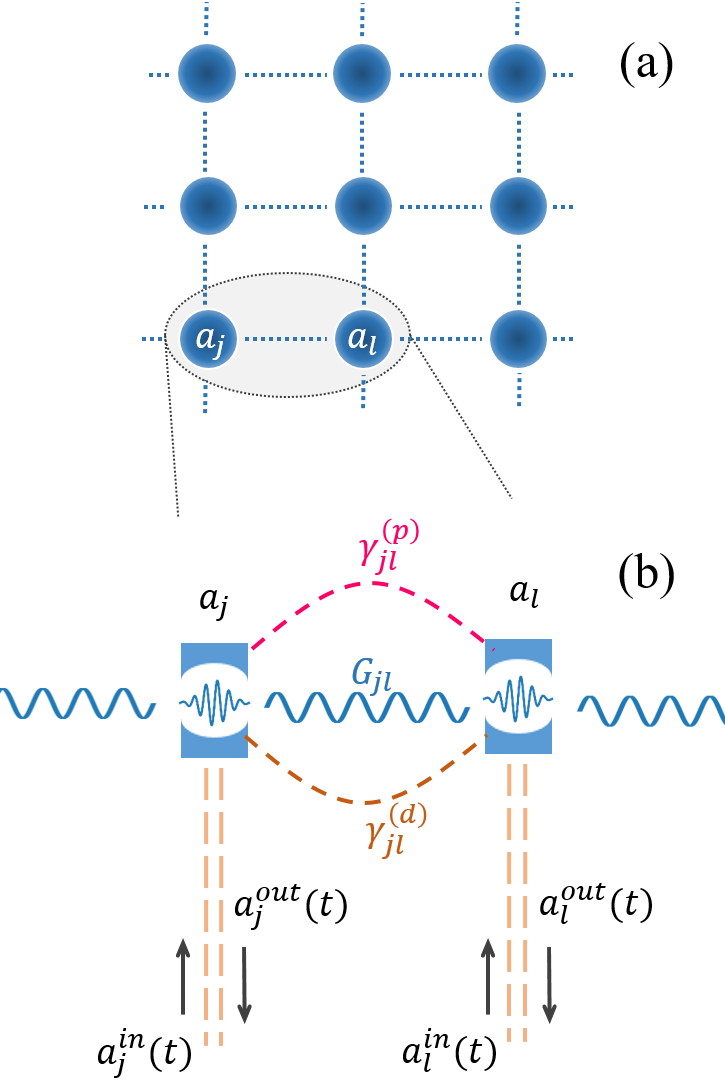}
	\caption{
	(a) Schematic representation of a photonic lattice consisting of a periodic system of coupled localized photonic modes. 
	(b) Two sites in a generic photonic lattice, together with the input/ouput fields and couplings described in the main text: $G_{jl}$ is a coherent, photon tunneling term,  whereas $\gamma^{(p)}_{jl}$ and $\gamma^{(d)}_{jl}$ are dissipative gain and loss couplings, respectively.}
	\label{twosites}
\end{figure}

The optical analog of an electronic lattice is a photonic lattice. By the latter term, we refer to any system that has a regular, periodic, structure, and whose physical properties can be understood in terms of localized photonic modes coupled by coherent or incoherent processes (see Fig. \ref{twosites} (a)).
Although in this work we are focused mostly on photonic quantum systems, our conclusions can be extended to vibronic lattices, where local modes are phononic, but lead to a similar phenomenology as their photonic counterparts.
Both photonic and vibronic lattices show distinctive features with respect to electronic crystals that complicate the characterization of topological effects. 
In particular, in those bosonic systems, gain and loss mechanisms are an intrinsic part of the dynamics, which induce decoherence and dissipation. The breaking of time-reversal invariance that is a typical ingredient of topological phases, together with dissipation, leads in photonic lattices to the occurrence of non-reciprocal photon transport \cite{Longhi17} and topological quantum fluctuations \cite{peano16prx}. Recently, it has been realized that these non-reciprocal effects may lead to directional amplification \cite{FernandezLorenzo19prl,wanjura2020topological}, in which a coherent input signal is exponentially amplified along the photonic chain. This phenomenon can be linked to the photonic lattice's topological properties \cite{peano16prx,FernandezLorenzo19prl,wanjura2020topological} and may have promising applications when combined, for instance, with current travelling-wave Josephson parametric amplifiers \cite{macklin_nearquantum-limited_2015,planat_photonic-crystal_2020,winkel_nondegenerate_2020,sivak_josephson_2020,renger_beyond_2020,malnou_three-wave_2020}. Part of the phenomenology of topological photonic lattices can be explained by directly applying topological insulator theory to non-Hermitian coupling matrices \cite{Gong18,shen18}. Nevertheless, non-Hermitian physics neglects essential aspects of the dissipative dynamics of a photonic lattice, such as quantum jumps. These limitations can be overcome with an input-output theory for topological bosonic systems that takes into account quantum noise and fluctuations, and thereby provides a full description of the dynamics as a many-body open quantum system.

In this work, we build on a formalism developed by some of us in Ref.~\cite{FernandezLorenzo19prl}, where photonic lattices were mapped to topological insulator Hamiltonians through the singular value decomposition (SVD) of the non-Hermitian linear coupling matrix.
This description showed the equivalence between the phenomenon of directional amplification and the occurrence of topologically non-trivial phases, which led to the concept of {\it topological amplification}. Here, we give a step forward and present a characterization of the quantum optical response of non-reciprocal photonic lattices in terms of the input-output formalism.

Our main results are the following: 
(i) We present an input-output theory of directional amplification in a photonic lattice that incorporates the connection to topological band theory from Ref.~\cite{FernandezLorenzo19prl}. 
The theory allows us to calculate the output signal and the output noise in the presence of a coherent input signal driving the lattice.
(ii) We identify topologically non-trivial phases in which directional amplification is linked to the existence of edge or surface states of an underlying topological insulator model. Our theory predicts different topological phases depending on the frequency of the input signal. 
(iii) We illustrate the theory in a one-dimensional array of cavities coupled by both coherent and dissipative terms, for which we derive fully analytical predictions.
(iv) For that example we show that, in the regime of topological amplification, the gain manifests an exponential growth as a function of the size, whereas its bandwidth decreases only with the inverse of the square root of the chain length. We prove that the added noise has a minimum value of $1$, slightly above the quantum limited value of $1/2$, and this is achieved for strong directional amplification regimes. We also show that in the regime of topological amplification, both the signal and noise undergo exponential amplification, however, the noise-to-signal ratio decreases as $\sim 1/\sqrt{N}$, being $N$ the number of sites. 
(v) Finally, we present numerical calculations that show that topological amplification is robust against disorder, up to a finite value of disorder strength, comparable to the photonic chain couplings strengths.

\section{Photonic lattice in the input-output formalism}
\subsection{Driven-dissipative dynamics and input-output operators}
We consider photonic lattices or arrays of coupled cavities described by Gaussian models, that is, without any photonic non-linearity involved. 
Local photonic modes have annihilation and creation operators $a_j$ and $a^\dagger_j$, respectively. 
Fig. ~\ref{twosites} displays a schematic representation of this system.

The coherent evolution of the photonic lattice is described by the Hamiltonian operator (units such that $\hbar = 1$),
\begin{equation}
H_c = \sum_j \omega_j a^\dagger_j a_j + \sum_{j,l} G_{j l} a^\dagger_j a_l ,
\end{equation}
where $\omega_j$ are the frequencies of each local photonic mode $j=1,\dots, N$, with $N$ the number of sites. In addition, $G_{jl}$ are complex tunneling terms satisfying $G_{jl}=G_{lj}^\ast$ and $G_{jj} = 0$. 

The incoherent dynamics includes pumping and dissipation, and it can be described within the master equation formalism as \cite{QuantumNoise},
\begin{align}
\dot{\rho} ={}& 
- i [H_{c}, \rho]  + \sum_j \kappa_j {\cal D}[a_j,a^\dagger_j]\left(\rho \right)
 \nonumber \\
+{}& \sum_{j,l} \gamma^{(d)}_{j l}  {\cal D}[a_j,a^\dagger_l]\left(\rho \right)
+ \sum_{j,l} \gamma^{(p)}_{j l}  {\cal D}[a_j^\dagger,a_l]\left(\rho \right).
\label{eq:master.equation}
\end{align}
Here, $\rho(t)$ is the density operator of the photonic lattice, and ${\cal D}[A,B] \left( \rho \right) = A \rho B - B A \rho/2 - \rho B A /2$ are generalized Lindblad superoperators. The second term in Eq.~(\ref{eq:master.equation}) describes local decay with rate $\kappa_j$ induced by the coupling to waveguides that transmit input and ouput fields [cf.~Fig.~\ref{twosites} (b)]. Finally, $\gamma^{\rm (d)}_{j l}$ and $\gamma^{\rm (p)}_{j l}$ are Hermitian positive semi-definite matrices that describe both local and collective photon losses and incoherent pumping, respectively. Non-diagonal dissipative terms ($j\neq l$) can be engineered and controlled by using additional degrees of freedom, like auxiliary intermediate modes that may lead to effective non-local dissipative couplings after tracing them out \cite{Hafezi12,Ruesink16,Shen16,bernier18pra}). 

Equivalently, we can describe the same driven-dissipative dynamics of the photonic lattice in the Heisenberg picture with the input-output formalism. To do so, it is first necessary to diagonalize the dissipative coupling terms in Eq.~(\ref{eq:master.equation}) and thereby write the master equation in Lindblad form \cite{QuantumNoise}. We thus define canonical collective modes
\begin{eqnarray}
\gamma^{(d)}_{jl} &=& \sum_{m} \bar{\gamma}_m^{(d)} T_{m j}^* T_{ml}, \nonumber \\
\gamma^{(p)}_{jl} &=& \sum_{m} \bar{\gamma}_m^{(p)} R_{m j}^* R_{ml},
\end{eqnarray}
with unitary matrices $T_{ml}$ and $R_{ml}$ and real eigenvalues $\bar{\gamma}^{(d)}_m$ and $\bar{\gamma}^{(p)}_m$, describing the collective rates for dissipation and 
incoherent pumping, respectively. Using the above decomposition, the master equation (\ref{eq:master.equation}) takes the Lindblad form,
\begin{align}
\dot{\rho} ={}& 
- i [H_{c}, \rho]  + \sum_j \kappa_j {\cal D}[a_j,a^\dagger_j]\left(\rho \right)
  \label{eq:master.equation_Lindblad}\\
{}&+ \sum_m \bar{\gamma}^{(d)}_m {\cal D}[d_m,d^\dagger_m]\left(\rho \right)
+ \sum_m \bar{\gamma}^{(p)}_m {\cal D}[p_m^\dag,p_m]\left(\rho \right).\nonumber
\end{align}
Here, the collective jump operators describing dissipation $d_m$ and pumping $p_m^\dagger$ are superpositions of local photonic modes given by
\begin{align}
d_m ={}& \sum_j T^*_{mj} a_j,\nonumber \\
p_m^\dagger ={}& \sum_j R_{mj}^* a_j^\dagger.\label{jumpOps}
\end{align}

The identification of these jump operators allows us to use the standard expressions of the input-output formalism to describe the open dynamics in the Heisenberg picture \cite{QuantumNoise}. In particular, the quantum Langevin equations of motion for the bosonic Heisenberg operators $a_j(t)$ read,
\begin{align}
\dot{a}_j ={}& i [H_c, a_j] -  \frac{\kappa_j}{2} a_j -\sqrt{\kappa_j}a_j^{\rm in}(t) +\sum_l \!\left(\frac{\gamma_{jl}^{(p)}}{2} - \frac{\gamma_{lj}^{(d)}}{2}\right) a_l\nonumber\\
+{}&\sum_m \sqrt{\bar{\gamma}_m^{(p)}}R_{mj}^\ast p_m^{in}{}^\dag(t)-\sum_m \sqrt{\bar{\gamma}_m^{(d)}}T_{mj}^\ast d_m^{in}(t).\label{eq:Langevin}
\end{align}
In this formalism, the decay of photonic modes is characterized by negative linear terms $\sim \kappa_c, \gamma_{jl}^{(d)}$, whereas pumping correspond to linear terms with positive rates $\sim \gamma_{jl}^{(p)}$. The normalization of the quantum state is ensured by the photonic input fields $a_j^{\rm in}(t)$, $d_m^{\rm in}(t)$, and $p_m^{\rm in}{}^\dag(t)$, one for each jump operator in Eq.~(\ref{eq:master.equation_Lindblad}). 
These bosonic operators satisfy canonical commutation relations, 
\begin{align}
    [a_j^{in}{(t)},a_{j'}^{in}{}^\dag(t')]={}&\delta_{jj'}\delta(t-t'),\\
    [d_m^{in}{(t)},d_{m'}^{in}{}^\dag(t')]={}&[p_m^{in}{(t)},p_{m'}^{in}{}^\dag(t')]=\delta_{mm'}\delta(t-t'),\nonumber
\end{align} 
and characterize the state of the photons entering on each of the input channels of the system. As sketched in Fig.~\ref{twosites} (b), operators $a^{in}_j(t)$ corresponds to photons on the input of lattice site $j=1,...,N$, whereas $d_m^{in}{(t)}$ and $p_m^{in}{}^\dag{(t)}$ are effective input operators from auxiliary modes that induce nonlocal decay and pumping, respectively. 

After the input photons interact with the lattice, these and other photons can leave the system through any output channel. In particular, photons leaving via the output of lattice site $j$ are described by $a_j^{out}(t)$, whereas photons leaving via the effective dissipation or pumping channels are modeled by $d^{out}_m(t)$ and $p^{out}_m(t)$  [cf.~Fig.~\ref{twosites} (b)]. 
These output operators are related to the input and system operators by the well-known input-output relations \cite{QuantumNoise},
\begin{eqnarray}
a^{out}_j(t) &=& a_j^{in}(t) + \sqrt{\kappa_j}  a_j(t),\\
d^{out}_m(t) &=& d_m^{in}(t) + \sqrt{\bar{\gamma}^{(d)}_m} d_m(t),\\
p^{out}_m(t) &=& p_m^{in}(t) + \sqrt{\bar{\gamma}^{(p)}_m} p^\dagger_m(t).
\label{eq:io}
\end{eqnarray}
and therefore, when combining with the equations of motion (\ref{eq:Langevin}), we can determine any correlation of the output photons. In particular, these equations will be very convenient to characterize the amplifying properties of the photonic lattice as shown below.

\subsection{Dynamics and output fields in frequency space}

In this subsection we solve for the dynamics and the output fields of the photonic lattice by moving to frequency space. 

We first rewrite the quantum Langevin equations (\ref{eq:Langevin}) very compactly as
\begin{eqnarray}
\dot{a}_j =  \sum_l H_{j l} a_l + \xi_j^{in}(t).
\label{eq:linear}
\end{eqnarray}
Here, we have introduced the non-Hermitian coupling matrix $H_{jl}$ as, 
\begin{equation}
H_{j l} 
= \Gamma_{j l}  - i \omega_j \delta_{j l} - i G_{j l},
\label{def.H}
\end{equation}
with $\Gamma_{j l}$ that combines all dissipative terms as,
\begin{equation}
\Gamma_{j l} 
= - \frac{\kappa_j}{2} \delta_{j l} + \frac{\gamma^{(p)}_{j l}}{2}   
- \frac{\gamma^{(d)}_{l j}}{2}.
\end{equation}
In addition, $\xi_j^{in}(t)$ in Eq.~\eqref{eq:linear} is the total input noise operator acting on lattice site $j$, and whose expression can be read from Eq.~\eqref{eq:Langevin},
\begin{align}
\xi^{in}_j(t) ={}& 
- \sqrt{\kappa_j} a_j^{in}(t) 
+ \sum_m \sqrt{\bar{\gamma}_m^{(p)}} R^*_{m j} {p^{in}_m}^\dagger(t)\nonumber \\
{}& - \sum_m \sqrt{\bar{\gamma}_m^{(d)}} T^*_{m j} d^{in}_m (t).\label{totalNoise}
\end{align}

Since the equations of motion (\ref{eq:linear}) are of first order and linear, they can solved exactly by Fourier transforming all operators as $\tilde{f}(\omega) = (2\pi)^{-1/2} \int_{-\infty}^{\infty} dt e^{i \omega t} f(t)$, and inverting the resulting algebraic matrix equations. Doing so, the frequency components of the lattice operators, $\tilde{a}_j(\omega)$, read 
\begin{equation}
\tilde{a}_j(\omega) = - \sum_l Q_{jl}(\omega) \tilde{\xi}^{in}_l(\omega),\label{solutionafreq}
\end{equation}
where $\tilde{\xi}^{in}_l(\omega)$ is the Fourier transform of Eq.~(\ref{totalNoise}) and the frequency-dependent response matrix, $Q(\omega)$, is given by the inverse of $H+i\omega\mathbb{1}$,~i.e.
\begin{equation}
Q_{jl}(\omega) = \left((H + i \omega \mathbb{1})^{-1}\right)_{jl}.
\label{eq:Q}
\end{equation}

The knowledge of $\tilde{a}_j(\omega)$ at all frequencies $\omega$ allows us to determine the dynamics of the photonic lattice exactly. Moreover, if we combine Eq.~(\ref{solutionafreq}) with the Fourier transform of the input-output relation in Eq. (\ref{eq:io}), we can solve for the frequency components of the output fields, $\tilde{a}_j^{out}(\omega)$, 
and describe photons leaving the lattice at site $j$. We find,
\begin{align}
\tilde{a}_j^{out}(\omega) ={}&  
\tilde{a}_j^{in}(\omega) - \sum_l \sqrt{\kappa_j} Q_{jl}(\omega) \tilde{\xi}^{in}_l(\omega)\\
={}&\sum_{l}Z_{jl}(\omega) \tilde{a}_l^{in}(\omega)
\nonumber \\
{}&
+\sum_{m,l} \sqrt{\kappa_j\bar{\gamma}_m^{(d)}} Q_{jl}(\omega) T^*_{m l} \tilde{d}^{in}_m(\omega) \nonumber\\
{}&-\sum_{m,l} \sqrt{\kappa_j\bar{\gamma}_m^{(p)}} Q_{jl}(\omega) R^*_{m l} \tilde{p}^{in}_m{}^\dag(\omega),\label{Fulloutputfield}
\end{align}
where we have defined the convenient function
\begin{equation}
Z_{jl}(\omega) = \delta_{jl} + \sqrt{\kappa_j \kappa_l}  Q_{jl}(\omega).
\end{equation}
Note that the output field $\tilde{a}_j^{out}(\omega)$ at site $j$ depends on the inputs at all channels $\tilde{a}_j^{in}(\omega)$, $\tilde{d}^{in}_m(\omega)$, and $\tilde{p}^{in}_m{}^\dag(\omega)$ because of the photonic hopping $G_{jl}$,
and collective incoherent terms $\Gamma_{jl}$ of the photonic lattice dynamics.

\subsection{Amplification of coherent fields: Gain, added noise, and noise-to-signal ratio}

In this subsection, we use the results above to describe the amplification of a nearly coherent input field that enters the photonic lattice at the first site, $j=1$, and then leaves at any site $j$. This scheme is independent of the dimension of the lattice and it can be easily extended to situations in which more than a site is coherently driven.

First we characterize the properties of the input signal. For convenience, we decompose the input field as
\begin{align}
a_j^{in}(t) = \langle a_j^{in}(t) \rangle + \delta a_j^{in}(t),\label{decompInput}    
\end{align}
where 
\begin{align}
\langle a_j^{in}(t) \rangle = \delta_{1j} \alpha e^{-i\omega_d t},
\label{coherentInput}
\end{align}
is the coherent field on the input port at site $j=1$ with amplitude $\alpha$ and frequency $\omega_d$. In addition, $\delta a_j^{in}(t)$ describes thermal fluctuations around the coherent amplitude such that $\langle\delta a_j^{in}(t)\rangle=0$. 
The noise associated to the input signal is well characterized in frequency space by the fluctuation correlations, 
\begin{align}
\langle \delta \tilde{a}^{in}_j(\omega)^\dagger  \delta \tilde{a}^{in}_j(\omega') \rangle = n^{in}(\omega)\delta_{j1}\delta(\omega - \omega'),    
\end{align} 
with $n^{in}(\omega)$ the number of input noise photons at frequency $\omega$.

Using the above description, the flux ${\cal F}_1^{in}$ of the input field can be expressed in terms of signal and noise components as
\begin{align}
{\cal F}_1^{in} = \langle a^{in \dagger}_1(t) a^{in}_1(t) \rangle = {\cal S}_1^{in} + {\cal N}_1^{in} ,
\end{align}
where the signal is the total input flux ${\cal S}_1^{in}=|\langle a_1^{in}(t)\rangle|^2=|\alpha|^2$, and the total input noise reads 
\begin{align}
    {\cal N}_1^{in} = \langle\delta a^{in}_1{}^\dag(t) \delta a^{in}_1(t)\rangle = \frac{1}{2\pi}\int d\omega\ n^{in}(\omega).
\end{align}

In the following we use the general expression for the output field $\tilde{a}_j^{out}(\omega)$ in Eq.~(\ref{Fulloutputfield}) to quantify how the input signal ${\cal S}_1^{in}$ and its noise ${\cal N}_1^{in}$ are amplified or modified by the photonic lattice. For this it is convenient to also decompose the output field in coherent and fluctuation terms as, $\tilde{a}_j^{out} (\omega) = 
\langle \tilde{a}_j^{out} (\omega) \rangle + \delta \tilde{a}_j^{out}  (\omega)$, with
\begin{align}
\langle \tilde{a}_j^{out} (\omega) \rangle \!={}&\!\! 
\sum_l Z_{jl}(\omega) \langle \tilde{a}_l^{in} (\omega) \rangle
\nonumber \\
\!={}&\! 
\sqrt{2 \pi} Z_{j1}(\omega)\alpha \delta(\omega-\omega_d) ,\label{coherentoutput} \\
\delta \tilde{a}_j^{out} (\omega)
\!={}&\!\!\sum_l Z_{jl}\delta \tilde{a}_l^{in}(\omega)\!+\!\sum_{l,m}\!\sqrt{\kappa_j\bar{\gamma}^{(d)}_m}T_{ml}^\ast Q_{jl}(\omega)\tilde{b}_m^{in}(\omega)\nonumber\\
{}&-\sum_{l,m}\!\sqrt{\kappa_j\bar{\gamma}^{(p)}_m}R_{ml}^\ast Q_{jl}(\omega)\tilde{d}_m^{in}{}^{\dagger}(\omega).\label{fluctuationoutput}
\end{align}
The photon output flux ${\cal F}_j^{out}(t)$ at site $j$ can then be calculated and decomposed in output signal and noise as
\begin{align}
{\cal F}_j^{out}(t)=\langle  a^{out}_j (t)^\dagger a^{out}_j (t) \rangle={\cal S}^{out}_j + {\cal N}^{out}_j.    
\end{align}
The output signal ${\cal S}_j^{out}$ reads
\begin{align}
{\cal S}_{j}^{out} ={}&|\langle a_j^{out}(t)\rangle|^2= G_j(\omega_d)|\alpha|^2,
\label{Sjfull}
\end{align}
where the gain $G_j(\omega_d)$ determines the amount of amplification of the input signal at frequency $\omega_d$. This gain is given by 
\begin{align}
    G_j(\omega) = |Z_{j1}(\omega)|^2.
\label{gain1}
\end{align}
For an output different than the input signal, $j\neq 1$, it simplifies to
\begin{equation}
G_{j\neq 1}(\omega)= \kappa_j\kappa_1|Q_{j1}(\omega)|^2.
\label{Sj}
\end{equation}
We see that the amplification amplitude and bandwidth can be fully characterized by evaluating $Q_{jl}(\omega)$ in Eq.~(\ref{eq:Q}), i.e.~the inverse of $H+i\omega\mathbb{1}$.

To study the noise contribution to the output field ${\cal N}_j^{out}$, we use Eq.~(\ref{fluctuationoutput}) and the properties of the input field to determine the correlation,
\begin{align}
    \langle \delta \tilde{a}^{out}_j(\omega)^\dagger  \delta \tilde{a}^{out}_j(\omega') \rangle= n_j^{out}(\omega) \delta(\omega - \omega').
    \label{outputnoisecalc}
\end{align}
The output noise is characterized by a density of photons $n^{out}_j(\omega)$ around frequency $\omega$, given by
\begin{align}
    n^{out}_j(\omega) = G_j(\omega)n^{in}(\omega) + n^{amp}_j(\omega).
\end{align}
The input noise $n^{in}(\omega)$ is also amplified by the gain factor $G_j(\omega)$ at frequency $\omega$. Importantly, it appears an extra term $n^{amp}_j(\omega)$ quantifying the noise added by the amplification process. 
Since we are assuming vacuum in all input ports except for the input field at $j=1$, we have
\begin{align}
    n^{amp}_j(\omega) =  \kappa_j  \sum_{l l'} Q_{j l}^\ast(\omega) Q_{j l'}(\omega) \gamma^{(p)}_{l'l}.
\label{Nw}
\end{align}
Therefore, the amplifier noise is also determined by the lattice response $Q_{jl}(\omega)$ together with the incoherent pump matrix $\gamma_{jl}^{(p)}$. 

Any linear non-parametric amplifier requires a component of incoherent pump $\gamma_{jl}^{(p)}\neq 0$ and coupling $\kappa_j$ to work. Consequently there will be always some extra noise $n^{amp}_j(\omega)$ added by the amplifier \cite{caves81prd}. Using Eq.~(\ref{outputnoisecalc}) we can calculate the total output noise, ${\cal N}_j^{out}$, which is given by the integral 
\begin{eqnarray}
{\cal N}_j^{out} = \langle\delta a^{out}_j{}^\dag(t) \delta a^{out}_j(t)\rangle =
\frac{1}{2\pi} \int  d \omega n_j^{out}(\omega).
\label{Nj}
\end{eqnarray}
Note that ${\cal N}^{out}_j$ describes a flux of incoherent photons leaving the photonic lattice at site $j$, even in the absence of any coherent input field.

We are now in position to discuss the quantum limits of the output noise added by the amplifying lattice and how the noise-to-signal ratio is affected by this. At the input, we have
\begin{align}
    \frac{ {\cal N}^{in} }{ {\cal S}^{in}}=\frac{1}{2\pi}\int d\omega \frac{n^{in}(\omega)}{|\alpha|^2},
\end{align}
an integral over the input noise, normalized by the signal strength $|\alpha|^2$. To see how the photonic lattice changes the noise-to-signal ratio, it is convenient to define the normalized added noise as
\begin{align}
n_j^{add}(\omega) = \frac{n^{amp}_j(\omega)}{G_j(\omega)},\label{addedNoise}  
\end{align}
which allows for a very direct comparison of the noise added by the amplifier and the input signal $|\alpha|^2$ assuming both are amplified by the gain factor $G_j(\omega)$. The noise-to-signal ratio \tr{\footnote{\tr{In the literature it is more common to define the signal-to-noise ratio, but in the present work we use the inverse quantity to discuss more precisely the noise error introduced by the amplifier.}}} at the output then reads
\begin{align}
    \frac{ {\cal N}_j^{out} }{ {\cal S} _j^{out}} = {}&\frac{1}{2\pi}\int d\omega \frac{G_j(\omega)}{G_j(\omega_d)}\left(\frac{n^{in}(\omega)}{|\alpha|^2}+\frac{n_j^{add}(\omega)}{|\alpha|^2}\right)\\
    \approx{}& \frac{{\cal N}^{in}_1}{{\cal S}^{in}_1}+\frac{1}{2\pi}\int d\omega \frac{n_j^{add}(\omega)}{|\alpha|^2}.
\end{align}
Due to the uncertainty relations, the added noise is bounded by $n_j^{add}(\omega)\geq 1/2$ \cite{caves81prd}. Therefore, the amplification of a quantum signal always increases the noise-to-signal-ratio. 

Nevertheless, we show below that the photonic lattice in the topological regime can behave as a nearly quantum limited amplifier, \tr{$n_j^{add}(\omega)\approx 1$}, and still amplify directionally. Moreover, we also show that the pre-factor of the signal to noise ratio reduces with the number of lattice sites as $\sim N^{-1/2}$, so that any input signal of flux $|\alpha|^2\gg {\cal N}^{in}_1 + \kappa_j/\sqrt{N}$ can be efficiently amplified with an exponential gain.

\section{From Directional amplification to Topological Band Theory} 
\subsection{Singular value decomposition and effective Hamiltonian}
The input-output formalism allows the quantification of the coherent output signal in terms of the response matrix $Q(\omega)$, i.e.~the inverse of the non-Hermitian matrix, $H + i \omega \one$. As pointed out in Ref.~\cite{FernandezLorenzo19prl}, $Q(\omega)$ may be computed using the singular eigensystem of $H + i\omega\one$ as, 
\begin{equation}
H + i \omega \one = U(\omega) S(\omega) V^\dagger(\omega),
\label{eq:svd1}
\end{equation}
where $S(\omega)$ is a diagonal matrix with non-negative real elements, and $U(\omega)$, $V(\omega)$ are unitary matrices. Using this decomposition, we write the inverse as
\begin{equation}
Q(\omega) = V(\omega) S(\omega)^{-1} U^\dagger(\omega).
\label{eq:svd2}
\end{equation}
To understand the link between Eqs.~(\ref{eq:svd1},\ref{eq:svd2}) and topological band theory, 
we introduce an auxiliary Hermitian matrix or effective Hamiltonian ${\cal H}(\omega)$, defined by
\begin{eqnarray}
{\cal H}(\omega)
&=&  
\begin{pmatrix}
\zero & H + i \omega \one  \\
H\tr{^\dag} - i \omega \one & \zero  
\end{pmatrix}
\nonumber \\
&=& (H + i \omega \one) \otimes \sigma^+ + (H^\dagger - i \omega \one) \otimes \sigma^- ,
\label{H}
\end{eqnarray}
where we have introduced ladder spin operators 
$\sigma^+$ and $\sigma^-$ acting on an auxiliary fictitious spin-$1/2$ spanned by states 
$\{ | \! \uparrow \rangle, |\!\downarrow \rangle  \}$. 

The importance of ${\cal H}(\omega)$ lies in the fact that {\it its eigensystem is the same as the singular value decomposition of $H + i \omega \one$}, namely
\begin{eqnarray}
{\cal H}(\omega) \begin{pmatrix}
U   \\
V   
\end{pmatrix}
&=& 
\begin{pmatrix}
U S   \\
V S  
\end{pmatrix} ,
\nonumber \\ 
\ 
{\cal H}(\omega) \begin{pmatrix}
U   \\
- V   
\end{pmatrix}
&=& 
- 
\begin{pmatrix}
U S   \\
- V S  
\end{pmatrix} .
\label{eq:svd3}
\end{eqnarray}
Thus the properties of the matrix $Q(\omega)$ can be analyzed by looking at properties of  ${\cal H}(\omega)$, with the advantage that ${\cal H}(\omega)$ is Hermitian and it can be characterized by TBT. This observation will allow us to relate the phenomenology of topological band insulators with the phenomenon of directional amplification.

Equation \eqref{eq:svd3} can be rewritten with the help of the fictitious spin-$1/2$ introduced in Eq.~\eqref{H}. 
To do so, we define the $n$th singular vectors 
$u^{n}(\omega)$ and $v^{n}(\omega)$ corresponding to the columns of $U(\omega)$ and $V(\omega)$, as 
$u^{n}_{j}(\omega) = U_{j n}(\omega)$, 
$v^{n}_{j}(\omega) = V_{j n}(\omega)$, which imply
\begin{eqnarray}
&& {\cal H}(\omega) 
\left( u^{n} \! \otimes \! | \! \uparrow \rangle 
\pm  
v^{n} \! \otimes \! | \! \downarrow \rangle \right) = 
\nonumber \\
&& \hspace{2cm} \pm s_n(\omega) 
\left( u^{n} \! \otimes \! | \! \uparrow \rangle 
\pm  
v^{n} \! \otimes \! | \! \downarrow \rangle \right).
\label{singularvectors}
\end{eqnarray}
We see that the eigenvalues of ${\cal H}(\omega)$ come in pairs, $\pm s_n(\omega)$, where the singular values are always $s_n(\omega) \geq 0$. 
The appearance of pairs of eigenvalues is due to the chiral symmetry 
\begin{equation}
(\mathbb{1}\otimes \sigma_z) {\cal H}(\omega) (\mathbb{1}\otimes \sigma_z) = - {\cal H}(\omega),
\label{chiral.symmetry}
\end{equation}
and Kramers degeneracy theorem.
We highlight that this chiral symmetry exists by the very definition of ${\cal H}(\omega)$ and it is independent of the underlying physical symmetries of the bosonic lattice. 
{\it Thus, chiral symmetry is never broken by any physical imperfection, on the contrary, it is a fundamental property of the dissipative lattice.}

\begin{figure}[t]
	\includegraphics[width=0.48\textwidth]{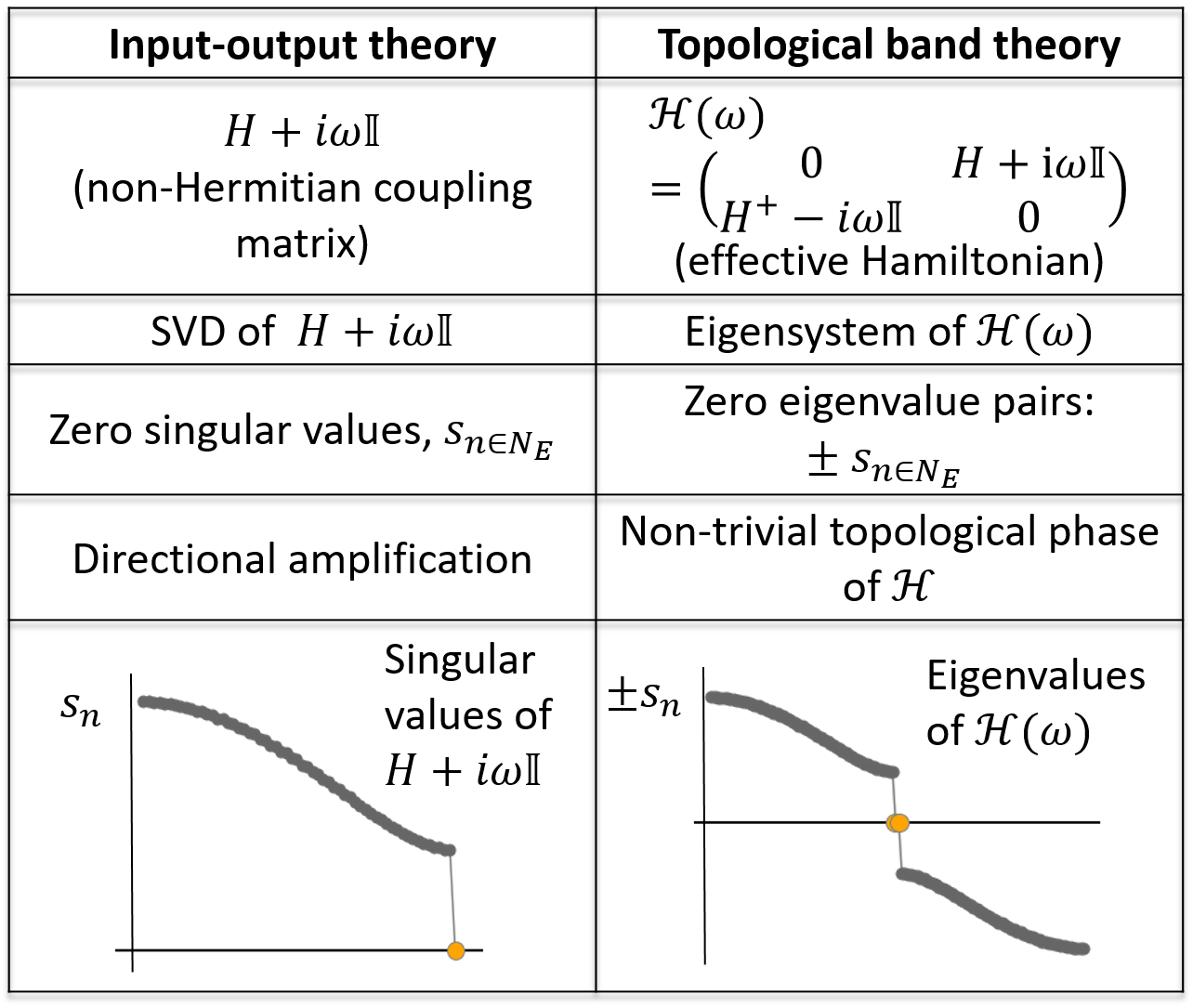}
	\caption{Mapping between non-Hermitian coupling matrices \eqref{def.H} and a chiral topological insulator Hamiltonian \eqref{H}. The mapping relies on expressing the SVD as the eigensystem of an effective extended Hamiltonian ${\cal H}(\omega)$. 
	The existence of zero-energy states in the topological insulator picture leads to the appearance of zero singular values of the coupling matrix and, thus, to the phenomenon of exponential directional amplification.}
	\label{fig:main_idea}
\end{figure}

We can explicitly write the singular value decomposition form of $Q_{jl}$, 
\begin{equation}
Q_{jl}(\omega) = \sum_n V_{j n}(\omega) (s_n(\omega))^{-1} \left( U_{l n}(\omega) \right)^* .
\label{Q.svd}
\end{equation}
Equation~\eqref{Q.svd} shows explicitly that the linear response of the photonic lattice is dominated by those singular vectors with small singular values. A comparison between the main aspects of the topological input-output and band theory is summarized in Fig.~\ref{fig:main_idea}.

\subsection{Edge singular vectors and directional amplification}
TBT predicts the existence of $N_E$ zero-energy eigenstates of ${\cal H}(\omega)$ in non-trivial topological phases (see for example Ref.~\cite{ryu02}). 
The latter are classified according to symmetry classes \cite{Schnyder08,ryu10,Bansil16,Chiu16}, 
which can be used to predict the existence of edge states from simple symmetry considerations.
The occurrence of zero-energy states in the spectrum of 
${\cal H}(\omega)$ implies the appearance of a set of $N_E$ zero-singular values in the SVD of 
$H + i\omega \one$.
Since the linear response of the photonic lattice is governed by small singular values, 
this implies that non-trivial topological properties of 
${\cal H}(\omega)$ have dramatic consequences in the steady-state of the photonic lattice.

We use the convention of sorting singular values in order of decreasing magnitude, 
such that zero-singular values in a non-trivial topological phase are the last $N_E$ singular values,  
$s_{n \in {\cal N}_E}$ with $n =$ $N - N_E-1, \dots, N$. 
Zero-singular values are separated by a gap from the bulk singular values, 
$s_{n = 1, \dots, N - N_E}$.
From TBT applied to ${\cal H}(\omega)$, we expect the emergence of right and left {\it edge} singular vectors, 
$u_j^{n \in {\cal N}_E}$ and $v_j^{n \in {\cal N}_E}$, whose amplitudes are localized 
at the edges of the lattice.  
In a finite-size system, zero energy states are not strictly zero, but typically they decrease exponentially with the size of the system. For example, in a one-dimensional non-trivial topological lattice, we expect that 
\begin{equation}
s_{n\in {\cal N}_E}(\omega) \propto e^{-N/\xi(\omega)}, 
\label{zero.sv}
\end{equation}
where $N$ is the number of sites and $\xi(\omega)$ is the edge-state localization length, which will in general depend on the frequency $\omega$. 

Based on the discussion above we can predict that non-trivial topological phases of ${\cal H}(\omega)$ 
have a dramatic effect on the matrix $Q(\omega)$. 
Imagine that ${\cal H}(\omega)$ is in a topologically non-trivial phase with a given number of edge states. 
The latter lead to zero singular values 
of $H + i \omega \one$, which dominate in the expression \eqref{eq:svd2}, such that
\begin{equation}
Q_{jl}(\omega) \approx \sum_{n \in {\cal N}_E} v_j^{n}(\omega)  \frac{1}{s_n(\omega)} 
{ u_j^{n}(\omega) }^* .
\end{equation}
Assume for simplicity that there is a single edge state, $n = N$, whose eigenvalue decreases exponentially with the system length, \tr{for which we obtain}  
\begin{equation}
Q_{jl}(\omega) \propto  v_j^{N}(\omega)  e^{N/\xi(\omega)} { u_j^{N}(\omega) }^* .
\end{equation}
We can simplify \tr{the above} expressions even further in a translationally invariant system, using the fact that $U$ and $V$ are related by spatial inversion. 
\tr{To show this,} we introduce the parity inversion operator, which can be defined by its action on an arbitrary vector $\psi$ as
\begin{equation}
\left( \Pi \psi \right)_j = \psi_{N - j + 1} . 
\end{equation}
\begin{figure}[t]
	\includegraphics[width=0.48\textwidth]{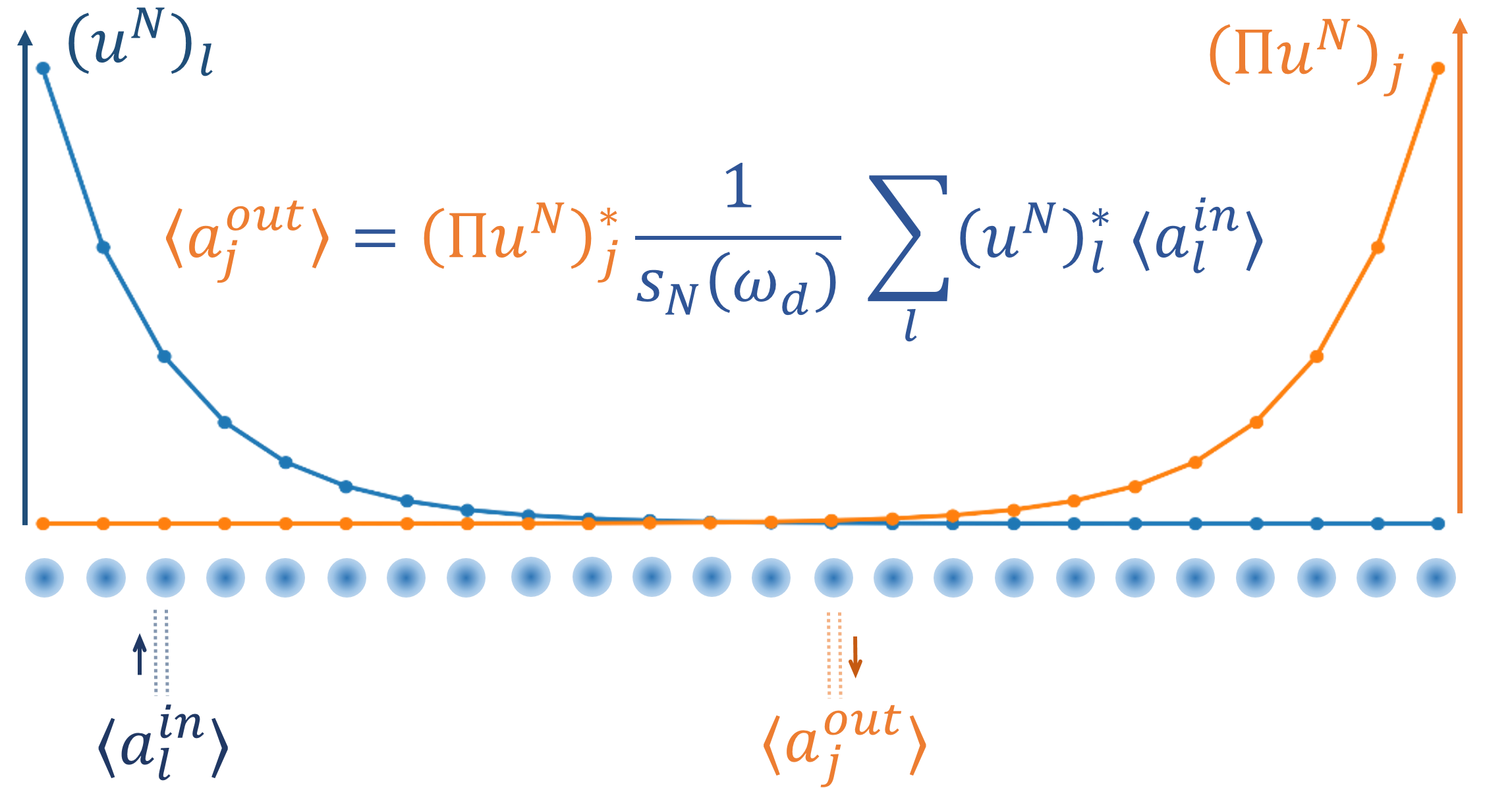}
	\caption{Representation of a 1D photonic chain that exemplifies the topological amplification process.
	A coherent input field drives the first lattice site $j = 1$ and this signal gets multiplied by the zero-singular vector $u^N$. The output signal at site $j=N$ is then proportional to the spatially inverted vector $\Pi u^N$ and it gets amplified by a factor $1/s_N(\omega_d)$.}
	\label{fig:scheme_io}
\end{figure}
In a translationally invariant system, this operator fulfills that 
\begin{equation}
\Pi H \Pi = H^{\rm T} .
\end{equation}
By substituting the SVD of $H + i \omega \one$ into the last equation, we can prove the relation 
\begin{equation}
V(\omega) = \Pi U^*(\omega), 
\end{equation}
which leads to the expression,
\begin{equation}
Q_{jl}(\omega) 
= - \sum_{n \in {\cal N}_E} \left( \Pi u^n(\omega) \right)^*_j 
\frac{1}{s_n(\omega)} 
 {u^n_l(\omega)}^*  .
\label{amplification}
\end{equation}

Using Eqs.~\eqref{amplification} and \eqref{zero.sv} in the expression for the lattice gain in Eq.~\eqref{Sj}, we see that \tr{it grows with the exponential} factor, 
\begin{equation}
G_j(\omega) \propto e^{N/\xi(\omega)}.
\end{equation}
\tr{Notice that $G_j(\omega)$} is also proportional to the overlap between the input signal and the singular edge vector $u_l^{(N)}(\omega)$
\tr{, but this is of order 1 when evaluated at the boundaries if $\xi(\omega) \approx 1$ , as shown below}. 
The coherent component of the output fields are distributed following the parity inverted singular edge-state vector $\Pi u_j^{(N)}(\omega)$.
This implies that amplification is a directional process triggered by a coherent drive in one of the system's edges and leading to large values of the field in the opposite edge \tr{[cf.~Fig.~\ref{fig:scheme_io}]}.

\subsection{Classification of Topological Phases}
The mapping from the non-Hermitian matrix, $H + i \omega \one$, to an effective Hamiltonian ${\cal H}(\omega)$ 
allows us to use the theoretical machinery of TBT \cite{Schnyder08,ryu10,Bansil16,Chiu16} and classify topological steady-states in translationally invariant lattices.  

Following TBT we consider periodic boundary solutions and express 
the matrix $H + i \omega \one$ in a plane-wave basis. We assume that the system is translationally invariant, and thus all the cavities or local bosonic modes have the same cavity frequency $\omega_0$. The Fourier transform is
\begin{equation}
( H + i \omega \one)_{\vec k} =  \Gamma_{\vec{k}} - i G_{\vec{k}} + i (\omega - \omega_0), 
\end{equation}
where $\Gamma_{\vec{k}}$ and $G_{\vec{k}}$ are real functions \tr{of the wavevector $\vec{k}$} due to the hermiticity of the coupling matrices. 
We find the following effective Hamiltonian,
\begin{eqnarray}
{\cal H}_{\vec{k}}(\omega) &=&  
\Gamma_{\vec{k}}  \sigma_x  + (G_{\vec{k}} + \omega_0 - \omega) \sigma_y  
\nonumber \\
&=& 
h_x(\vec{k}) \sigma_x + (h_y(\vec{k}) - \omega) \sigma_y.
\label{H.eff}
\end{eqnarray}
\tr{Here,} the two-dimensional vector 
\begin{equation}
  \vec{h}(\vec{k},\omega) = ( h_x(\vec{k}), h_y(\vec{k}) - \omega ), 
\end{equation}
defines a trajectory in 
$\vec{k}$-space that is linked to the appearance of topological invariants. We find that the topological phase depends on the frequency $\omega$, and thus, different spectral components of input/output field \tr{can be in topologically distinct regimes.} 

Let us consider, for example, a one-dimensional lattice in which case $\vec{k}$ becomes a scalar quantity $k$. 
Here we can define a winding-number $\nu(\omega)$ as the number of times that the vector $\vec{h}(k,\omega)$ encompasses the zero as $k$ goes from $0$ to $2 \pi$. Note that this winding number depends on the frequency $\omega$ of the fields. The topological amplification mechanism acts very differently on the coherent and incoherent ouput components. Let us discuss each case separately:

{\it Output signal (coherent).-} 
The coherent \tr{output} component ${\cal S}_j^{out}$ at a given site depends on the value of the \tr{response} matrix $Q_{jl}(\omega)$ at the frequency of the incoming coherent \tr{field, i.e.}~$\omega = \omega_d$. 
Thus, any property such as exponential amplification or existence of edge-states will be solely governed by the winding number value at that frequency, namely $\nu(\omega_d)$.

{\it Output noise (incoherent).-} 
On the contrary, the \tr{noise} component ${\cal N}_j^{out}$ is generated by the amplification of incoherently generated photons at any frequency $\omega$. Every $\omega$ has its own winding number. 
Those frequencies for which $\nu(\omega) = 1$ correspond to a topologically non-trivial phase and, thus, they dominate the incoherent emission process.

\begin{figure}[t]
	\includegraphics[width=0.48\textwidth]{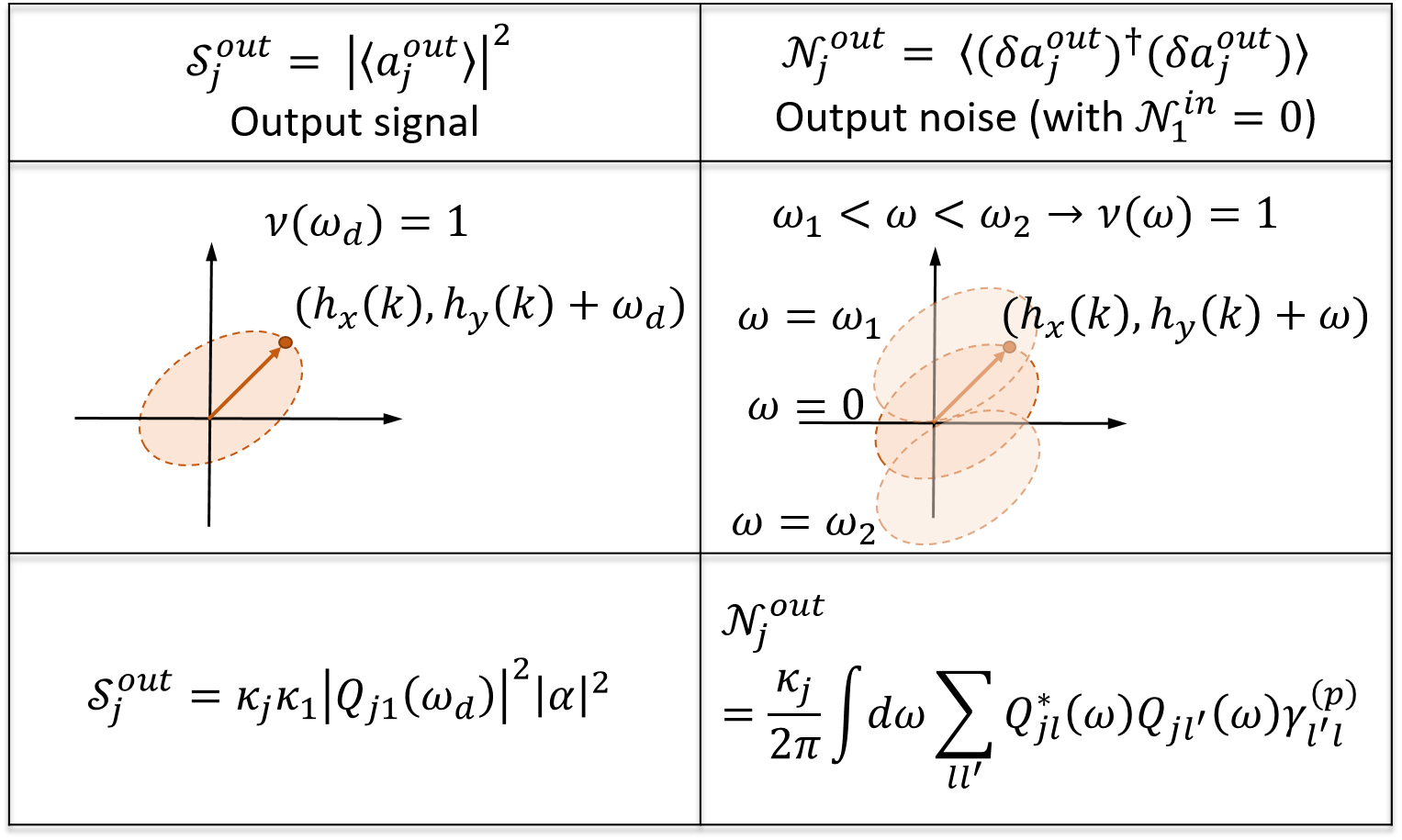}
	\caption{Different definitions of a topological index depending on whether we focus on the (coherent) output signal ${\cal S}^{out}_j$ or the (incoherent) output noise flux ${\cal N}_j$. 
	Left: The amplification of a coherent input signal (in this example, at lattice site $j = 1$) is determined by the matrix $Q(\omega)$, evaluated at the frequency of the coherent drive, $\omega = \omega_d$. Following the connection with the SVD described in the text, topological amplification occurs at non-zero values of the winding number $\nu(\omega_d)$. 
	Right: The output noise, on the contrary, receives contributions from the propagation of incoherently generated photons at any frequency $\omega$. 
	${\cal N}^{out}_j$ (assuming no significant noisy component in the input signal, ${\cal N}^{in}_1 = 0$) is dominated by those frequencies for which $\nu(\omega) \neq 0$ and, thus, topological amplification occurs.}
	\label{fig:fig2}
\end{figure}

The difference between those two situations is summarized in Fig.~\ref{fig:fig2}.

\subsection{Classification of topological amplifiers following the ten-fold way}

The classification of topological phases in TBT  relies on discrete symmetry operators such as \tr{time reversal $T$ and charge conjugation $C$. These can be written as}  
\begin{eqnarray}
T &=& U_T K , \nonumber \\
C &=& U_C K ,
\end{eqnarray}
where $U_T$, $U_C$ are unitary matrices and $K$ is the complex conjugation operator
($K^2 = 1$, $K i K = -i$). 
Time reversal and charge conjugation operators must fulfill the conditions $T^2 = \pm 1$ and $C^2 = \pm 1$, leading to the following restriction for the unitary matrices,
\begin{equation}
U_T U_T^* = \pm \mathbb{1}, \ \ U_C U_C^* = \pm \mathbb{1} .
\label{restriction}
\end{equation} 
Finally, $T$ and $C$, are related to the chiral symmetry $S$ as 
\begin{equation}
T C = S. 
\label{TC.condition}
\end{equation}
Since $S$ is defined by Eq.~\eqref{chiral.symmetry}, the relation
\begin{equation}
U_T U_C^* \propto \sigma_z ,    
\end{equation}
must be fulfilled so that the symmetry definitions are consistent.

\tr{In a translationally invariant system and going to Fourier space,} time-reversal and/or charge conjugation symmetries are fulfilled if there exist unitary matrices $U_T$, $U_C$, such that 
\begin{equation}
T {\cal H}_{\vec{k}}(\omega) T^{-1} = {\cal H}_{-\vec{k}}(\omega),
\label{time}
\end{equation}
and/or
\begin{equation}
C {\cal H}_{\vec{k}}(\omega) C^{-1} = -{\cal H}_{-\vec{k}}(\omega),
\label{charge}
\end{equation}
respectively. 

In appendix \ref{app:1} we show that symmetry classes AIII, BDI, CI and DIII are the only ones arising in the effective Hamiltonian representation of photonic lattices.

\section{Stability of the photonic lattice}\label{section:stability}
A non-trivial feature of an amplifier device is stability, which is determined by 
the eigensystem of the non-Hermitian matrix $H$,
\begin{equation}
H = B \Lambda B^{-1},
\end{equation}
where $\Lambda_{n m} = \delta_{n,m} \lambda_n$ is a diagonal matrix of eigenvalues and $B$ is a matrix of eigenvectors, which is generally not unitary. 
Following Eq.~\eqref{eq:linear}, we find that the system is stable only if
\begin{equation}
\mathfrak{Re} \left( \lambda_n \right) < 0,  \ \forall n,
\label{eq:stability}
\end{equation}
since -otherwise- any fluctuation gets exponentially amplified in time. 

We can use the master equation \eqref{eq:master.equation} to calculate the number of incoherent photons in the photonic lattice. This complementary approach must give the same prediction as the input-output formalism. \tr{To check this,} we define first the fluctuation operators in the lattice as,
\begin{equation}
\delta a_j  = a_j - \langle a_j \rangle.
\end{equation}
In absence of a coherent input field \tr{this would simplify to} $\langle a_j \rangle = 0$, and $\delta a_j = a_j$.
We also define the correlation matrix
${M}_{j l} =   \langle \delta a^\dagger_j \delta a_l \rangle$, whose time evolution 
can be directly obtained from the master equation \eqref{eq:master.equation},
\begin{eqnarray}
\dot{M}_{j l}  = \sum_{j'} H^*_{j j'} M_{j' l} + 
\sum_{l'} H_{l l'} M_{j l'} + \gamma^{(p)}_{l j} .
\label{eq:2point}
\end{eqnarray}

The correlation matrix takes a steady-state value $M_{jl}^{ss} = M_{jl}(t \to \infty)$), which can be calculated using the eigensystem of $H$ as,
\begin{align}
M_{j l}^{\rm ss} = & 
\nonumber \\
\sum_{n,m,j',l'} & B^*_{j n} B_{l m}
\frac{-1}{\lambda^*_n + \lambda_m} 
\left(B^{-1} \right)^*_{n j'} \left(B^{-1} \right)_{m l'} \gamma^{(p)}_{j'l'}.
\end{align}
The latter expression can be rewritten in integral form as,
\begin{equation}
M^{ss}_{jl} = \frac{1}{2\pi} \int d\omega \sum_{j',l'} Q_{jj'}^* (\omega) Q_{ll'} (\omega) 
 \gamma^{(p)}_{l' j'} .
\label{MssQ}
\end{equation}
\tr{This expression is obtained using} the definition of the matrix $Q(\omega)$ in Eq.~\eqref{eq:Q}, expressing
$(H + i \omega \one)^{-1}$ in terms of the eigensystem of $H$, and carrying out the intergration over $\omega$. 
Equation~\eqref{MssQ} is consistent with the result obtained with the input-output formalism 
for the output noise, since if fulfills that
\begin{equation}
    {\cal N}^{out}_j=\kappa_j M_{jj}^{ss},
\end{equation}
which agrees with the result from Eqs.~(\ref{Nw},\ref{Nj}) if we assume that ${\cal N}^{in}_1=0$.

\section{One-dimensional example: non-reciprocal photonic chain}
In this section we apply our input-output formalism to the particular case of a non-reciprocal photonic chain, strongly related to the Hatano-Nelson model \cite{Hatano96,Longhi15}. 
This model has the advantage that it leads to analytic results over a wide range of parameters and it can be used to test the validity of our predicitions, as well as to explore regimes of topological amplification.
\label{section:example}
\subsection{Definition of the model}
\begin{figure}[t]
  \includegraphics[width=0.48\textwidth]{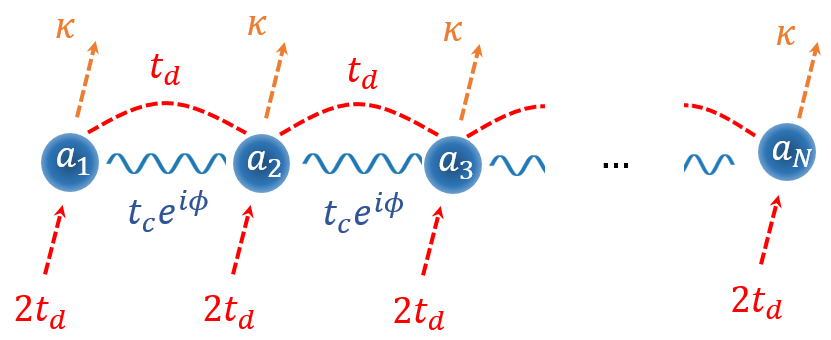}
  \caption{Schematic representation of the non-reciprocal photonic lattice studied in Section \ref{section:example}.} 
  \label{HatanoNelson} 
\end{figure}

We consider an array of cavities with nearest-neighbour dissipative and coherent couplings represented schematically in Fig. \ref{HatanoNelson}, leading to the coupling matrices \cite{FernandezLorenzo19prl},

\begin{eqnarray}
G_{j l} &=&  \ t_{c} e^{i \phi} \delta_{l,j+1} 
+ \ t_{c} e^{-i \phi} \delta_{l,j-1} ,  \nonumber \\
\Gamma_{j l} &=&  - \frac{\kappa}{2} \delta_{j l} +  2 t_d \delta_{j l} + t_{d} \delta_{l,j+1} + t_{d} \delta_{l,j-1} .
\label{1D.couplings}
\end{eqnarray}
Here, $t_c e^{i \phi}$ is a complex photon tunneling term with a phase $\phi$, which is required to break the time-reversal invariance of the system. 
This kind of coupling - with $\phi = 0$ - appears naturally in photonic setups, for example by connecting two microwave cavities in circuit QED \cite{houck12natphys}. Complex tunneling terms with $\phi \neq 0$ can be induced, for example, by means of Floquet engineering with periodic drivings \cite{Peropadre13, FernandezLorenzo19prl}.
In addition, $t_{\rm d}$ is a dissipative coupling, which can be induced, for example, by coupling to an auxiliary lossy cavity  \cite{Metelmann17,FernandezLorenzo19prl}. 
Since the coupling $t_d$ is obtained after tracing out a common dissipative reservoir, the natural assumption is to consider additional diagonal terms $2 t_d$, as can be shown by the explicit derivation of this model, see for example our previous reference \cite{FernandezLorenzo19prl}. 
Finally, $\kappa$ is the photon decay out of the photonic sites of the chain. To connect with the notation employed in the previous section, we can explicitly write the matrix
\begin{eqnarray}
\gamma_{jl}^{(p)} &=& 4 t_d \delta_{j,l} + 2 t_d \delta_{j,l+1} + 2 t_d \delta_{j,l-1},
\label{dissipative.couplings}
\end{eqnarray}

Although the Eq. (\ref{1D.couplings}) is a perfectly valid description of our non-reciprocal chain, the topological properties of the system are more intuitively understood in terms of the parameter
\begin{equation}
\gamma_p = 4 t_d - \kappa/2.
\label{dissipative.couplings}
\end{equation}
Here, $\gamma_p$ is the rate of incoherent pumping of photons into the photonic chain. 
The model defined by Eqs. (\ref{1D.couplings}) is not only the simplest example of a topological amplifier, but it can also be implemented by means of Floquet engineering of a photonic ladder in a superconducting circuit (see \cite{FernandezLorenzo19prl}).

The effective Hamiltonian in the plane-wave basis is
\begin{align}
{\cal H}_k(\omega) 
= & \left(\gamma_{\rm p} \!  - \! 2 t_\dd \! + \! 2 t_\dd \cos(k) \right) \sigma_x 
\nonumber \\
 & +  
\left( 2 t_\cc \cos(k \! + \! \phi)  + \omega_0 - \omega \right)\sigma_y .
\label{eq:H1D}
\end{align}
The cases $\phi = 0, \pi$ belong to the CI symmetry class of the ten-fold way classification.
The generic case with $\phi \neq 0$ belongs to the AIII class and here we can expect non-trivial topological phases to appear \cite{Chiu16}.

To characterize the properties of Hamiltonian \eqref{eq:H1D} we use the winding number $\nu(\omega)$ as a topological invariant \cite{asboth16}. 
\tr{This number $\nu(\omega)$ correspond to the} times that the circle formed by $\tr{\vec{h}(k)=}(h_x(k), h_y(k)\tr{-}\omega)$ 
encompasses the origin, $(0,0)$, with
\begin{eqnarray}
h_x(k) &=& \gamma_p - 2 t_d + 2 t_d \cos(k), \nonumber \\
h_y(k) - \omega &=& 2 t_c \cos(k+\phi) + \omega_0 - \omega .
\label{h.vector}
\end{eqnarray}
We calculate first the values $k_+, k_-$ at which 
$h_y(k_{\pm}) - \omega = 0$, \tr{which leads us to}  
$k_\pm = -\phi \pm \arccos((\omega-\omega_0)/(2 t_c))$. 
Then we calculate $h_x(k_\pm)$ and check whether $0 \in [h_x(k_-),h_x(k_+)]$, which is the condition for $\nu(\omega) = 1$.

\begin{figure}[t]
	\includegraphics[width=0.50\textwidth]{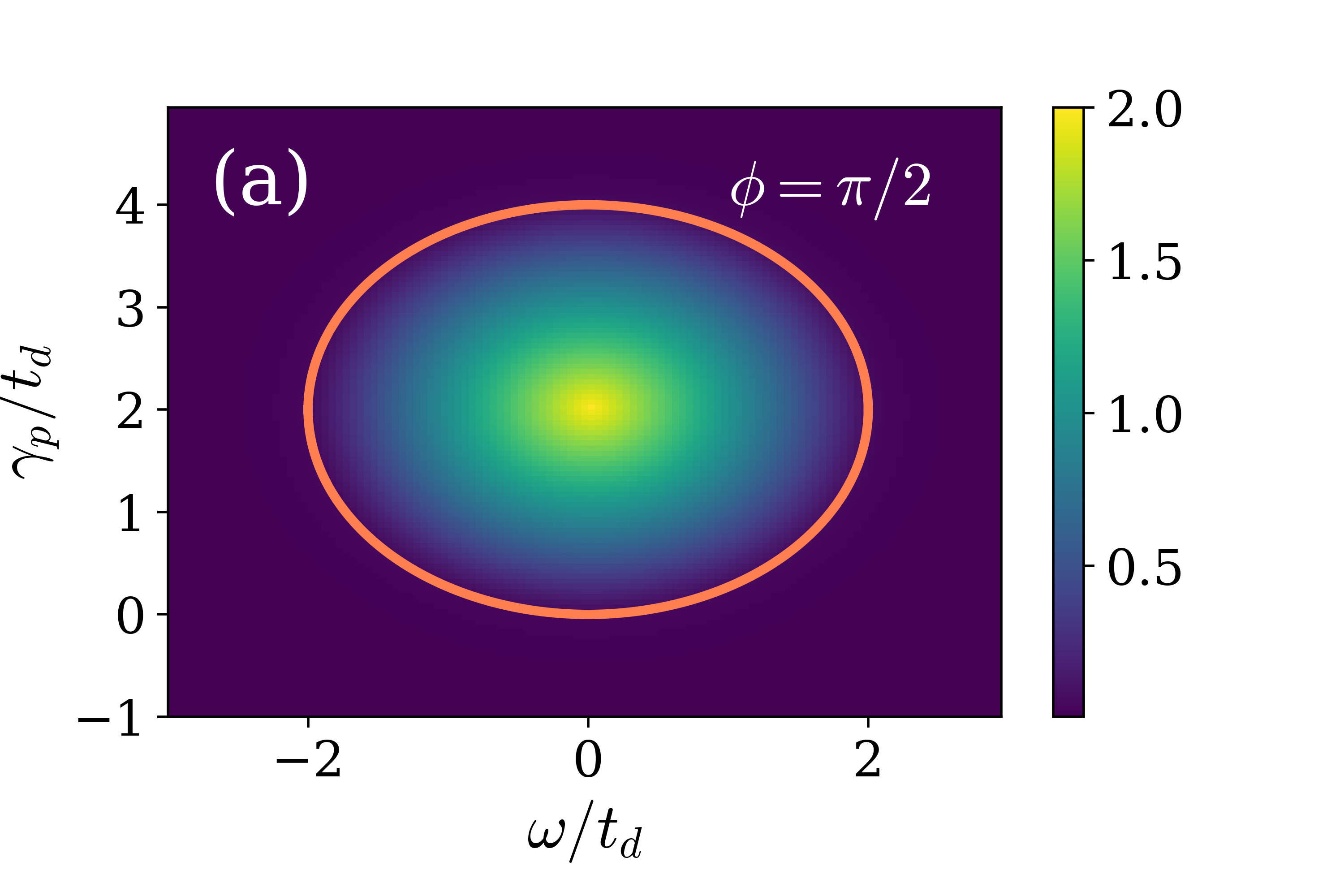}
	\includegraphics[width=0.50\textwidth]{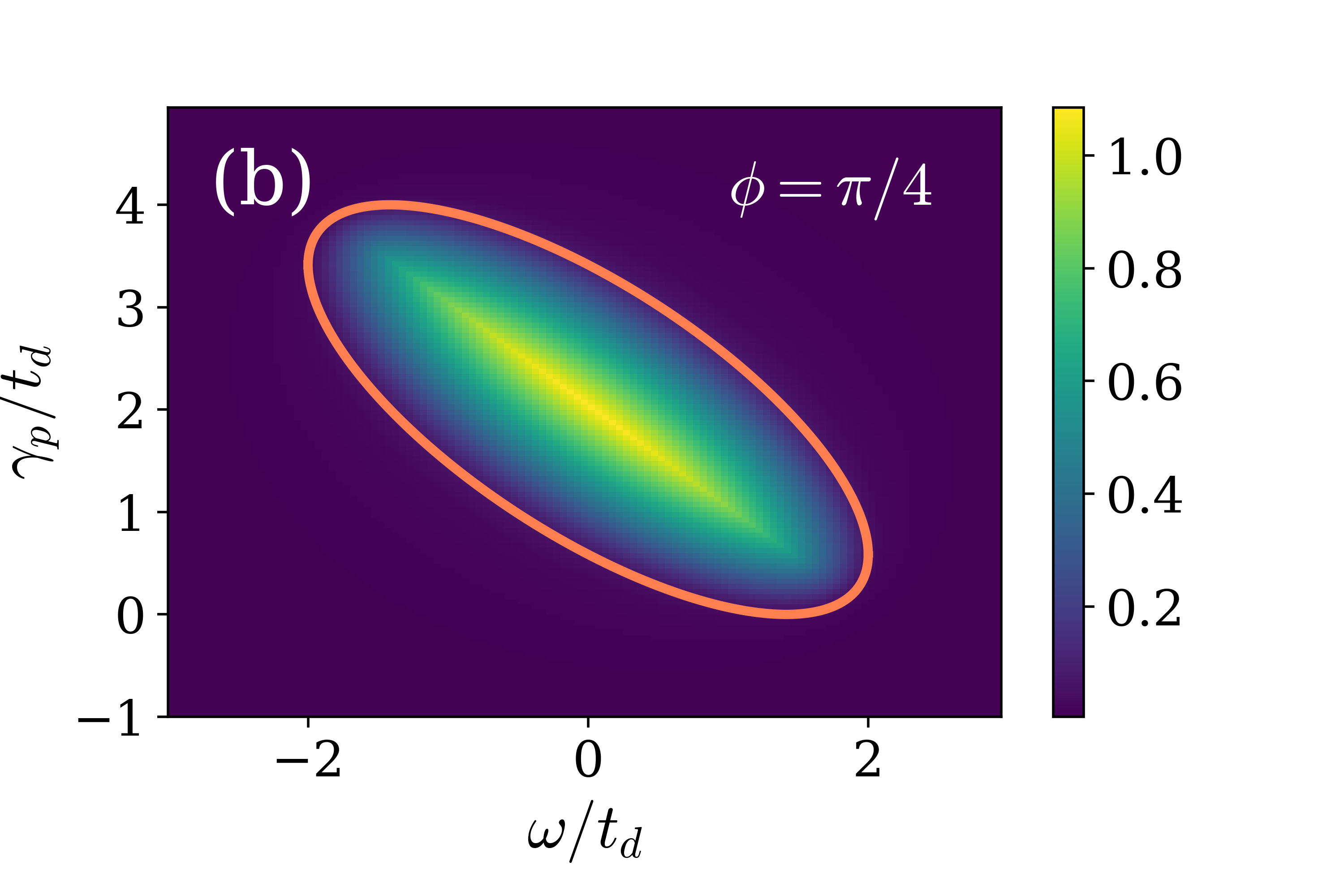}
	\caption{Singular value gap in units of $t_d$, $\Delta_g/t_d = (s_N - s_{N-1})/t_d$, 
	calculated numerically for the nearest-neighbor photonic chain 
	in Eq.~\eqref{1D.couplings}, with $N = 100$ sites and $t_d = t_c$, $\omega_0 = 0$. 
	We consider different values of $\omega$ and the net photon increase rate, $\gamma_p$. 
	The orange curve represents the boundary between the topological non-trivial and trivial phases as determined by the condition Eq. \eqref{winding_condition}. Our winding number criterion correctly identifies the topologically nontrivial regions as regions with a non-zero energy gap between zero singular value states and the bulk.}
	\label{singularvaluegap}
\end{figure}

Following the procedure above we find that the conditions for $\nu(\omega) = 1$ finally read
\begin{eqnarray}
\gamma_p^{-} &<& \gamma_p < \gamma_p^{+},  \nonumber \\
(\omega-\omega_0) &<& 2 t_c ,\label{conditions_topological}
\end{eqnarray}
where we have defined the critical photon pumping rates
\begin{align}
\gamma_p^{\pm} = 2 t_d - (\omega-\omega_0)&\frac{t_d}{t_c} \cos(\phi)\pm 2 t_d \sin\phi \sqrt{1-\frac{(\omega-\omega_0)^2}{(2 t_c)^2} } .
\label{winding_condition}
\end{align}
We have checked numerically the the above conditions agree with the numerical calculation of the singular value gap $\Delta_g = s_N - s_{N-1}$, see Fig. \ref{singularvaluegap}

From the results above it can be easily checked that the cases 
$\phi = 0$ and/or $\gamma_\pp = 0$ only have topologically trivial solutions. Thus, \tr{this} model requires complex photon tunneling couplings together with incoherent photon pump \tr{to manifest edge-states and topologically non-trivial phases}.

\subsection{Analytical results in the SSH limit}
We focus now on the case $\phi = \pi/2$ and $t_d = t_c$. 
This will allow us to simplify the discussion and range of parameters. 
The effective Hamiltonian is reduced to
\begin{eqnarray}
{\cal H}_k(\omega) &=& (\gamma_p - 2t_d + 2t_d\cos(k))\sigma_x 
\nonumber \\
&+& \left( (\omega - \omega_0) + 2 t_d \sin(k) \right) \sigma_y .
\label{Hk}
\end{eqnarray}
In the resonant case, $\omega = \omega_0$, the effective band Hamiltonian ${\cal H}(\omega)$ can be directly mapped into the celebrated  Su-Schriefer-Heeger (SSH) model \cite{Heeger88}, whose Hamiltonian takes the matrix form,
\begin{equation}
H_k^{\rm SSH} = \left( J_1 + J_2 \cos(k) \right) \sigma_x - J_2 \sin(k) \sigma_y .
\label{Hk.ssh}
\end{equation}
where $J_1$ and $J_2$ are the alternating hopping constants of the SSH chain. By comparing expressions \eqref{Hk} and \eqref{Hk.ssh}, we see that, apart from a trivial sign in $J_2$, the difference between the two models is the additional constant term, $\omega - \omega_0$, added in the prefactor of the $\sigma_y$ term in ${\cal H}_k(\omega)$. 
However, a rotation in the $x-y$ plane can bring \eqref{Hk} into the standard form of the SSH model,
\begin{align}
\bar{{\cal H}}_k(\omega) & = e^{i \theta \sigma_z/2} {\cal H}_k(\omega) e^{- i \theta \sigma_z/2} = \nonumber \\ 
& \left( r(\omega) + 2 t_d \cos(k) \right) \sigma_x + 2 t_d \sin(k) \sigma_y   ,
\end{align}
where $r(\omega)$ is the distance between the center of the circle spanned by the vector $\vec{h}_k=(h_x(k), h_y(k)-\omega)$ and the origin,
\begin{equation}
r(\omega) = \sqrt{(\omega-\omega_0)^2 + (\gamma_p - 2 t_d)^2}.
\end{equation}
The eigenvectors of ${\cal H}_k(\omega)$ and ${\bar {\cal H}}_k(\omega)$ are the same up to a phase between the singular vectors $u^n$ and $v^n$ (see Eq. \ref{singularvectors}).
Therefore, the expressions for the energy or localization lengths of zero-energy modes in the original SSH model can be generalized to our photonic dissipative chain by replacing,
\begin{align}
J_2 &\to 2 t_d \nonumber \\   
J_1 &\to r(\omega).
\label{equivalence.ssh}
\end{align}

Eq. ~\eqref{winding_condition}, together with conditions $\phi = \pi/2$, $t_d = t_c$, imply that non-trivial topological phases exist as long as  
\begin{equation}
r(\omega) < 2 t_d .
\label{cond.top}
\end{equation}
We can find analytically expressions for the edge state wave-functions by using the results obtained for the SSH model (see Ref.~\cite{Venuti07} for an analytical derivation of these expressions), together with the \tr{identification} \eqref{equivalence.ssh}. \tr{As a result,} the edge-state localization length $\xi(\omega)$ \tr{is given by}
\begin{equation} 
\frac{1}{\xi(\omega)} = \ln \left( \frac{2 t_d}{r(\omega)} \right) .
\label{def.xi}
\end{equation}
In the following, we assume the limit $N \gg \xi(\omega)$, such that finite-size effects can be neglected in what concerns the form of the edge singular vector $u_j^{(N)}(\omega)$. \tr{For $j=1,\dots,N$,} this takes the form 
\begin{eqnarray}
u_j^{(N)}(\omega) 
= \frac{1}{\sqrt{\rho(\omega)}} e^{-(j-1)/\xi(\omega)},
\end{eqnarray}
\tr{where} the normalization constant is
\begin{equation}
\rho(\omega) = 1/(1-e^{-2/\xi(\omega)}).
\end{equation}
The singular value of the edge mode can also be \tr{explicitly} calculated,
\begin{eqnarray}
s_N(\omega) &=& s_0(\omega) e^{-N/\xi(\omega)},\qquad \text{with,} \\
s_0(\omega) &=& 2 t_d \left( 1 - (r(\omega)/(2 t_d))^2 \right) .
\end{eqnarray}
The exponential part of $s_N(\omega)$ leads to exponential amplification \tr{as it will be illustrated below}.

The \tr{above} expressions, together with Eq.~\eqref{amplification}, allow us to calculate explicitly the matrix $Q_{jl}(\omega)$  in the topologically nontrivial regime,
\begin{equation}
Q_{jl}(\omega) = 
\frac{-1}{\rho(\omega) s_0(\omega)} e^{N/\xi(\omega)}   
e^{-(N-j)/\xi(\omega)} e^{-(l-1)/\xi(\omega)} .
\label{Q.edge}
\end{equation}
This last equation \tr{contains all} main results for our one-dimensional \tr{photonic lattice model}, and it can be used to fully characterize its amplification properties.

\subsection{Output signal and gain}
\begin{figure}[t]
	\includegraphics[width=0.45\textwidth]{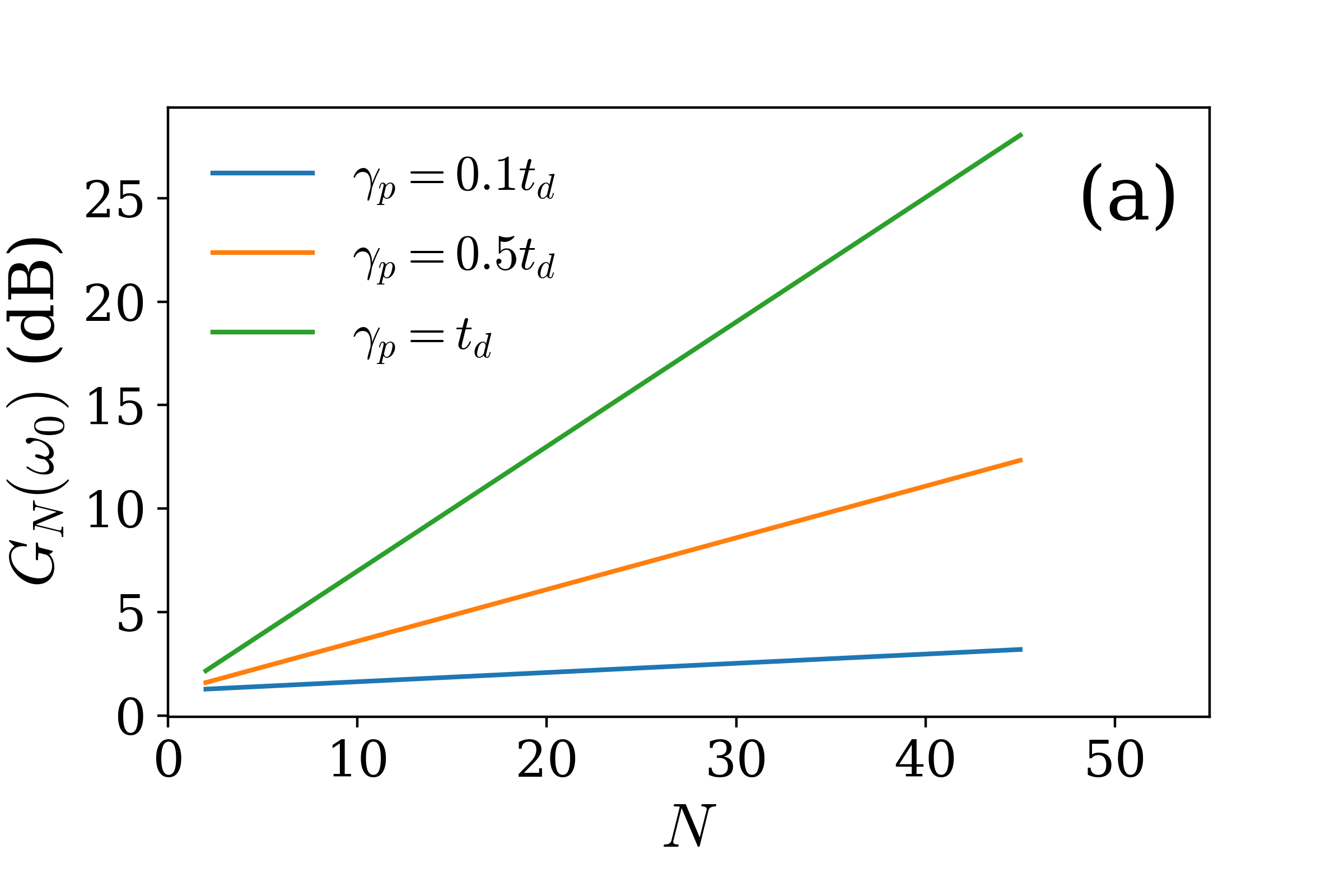}
	\includegraphics[width=0.45\textwidth]{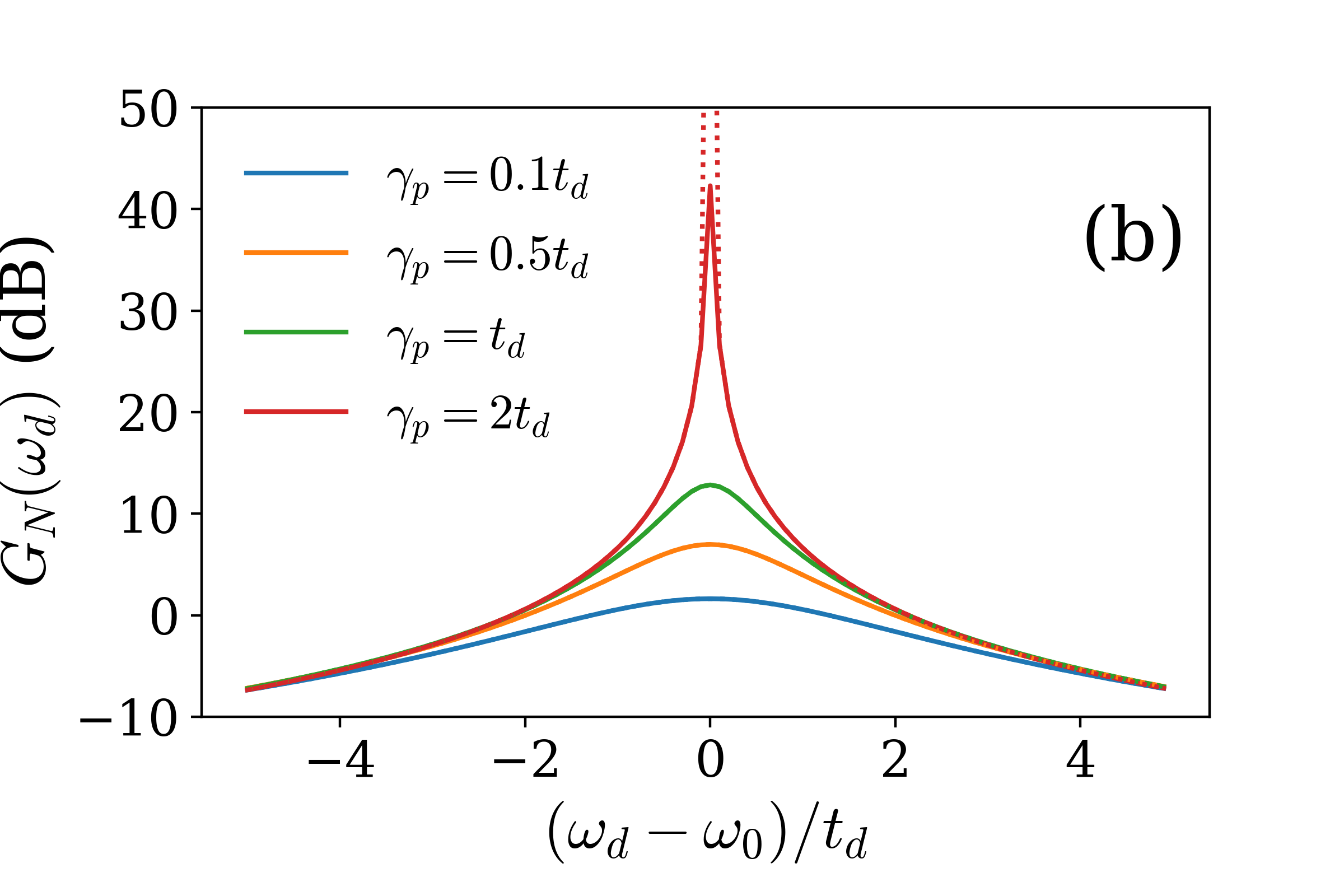}
  \caption{
  Gain at the last site, $j = N$, of the photonic chain defined by Eq. \eqref{1D.couplings}, with $t_d = t_c$, $\phi = \pi/2$, \tr{$\omega_d=\omega_0$,} and different values of the net photon pumping rate $\gamma_p$. The gain is expressed in decibels (dB), so that $10 \log_{10} G_N$ is plotted.
  Continuous lines are calculated by numerically evaluating the matrix $Q_{jl}(\omega)$ and then using Eq. \eqref{Sj}. 
  Dashed lines are calculated  and by using the analytical SSH solution in Eq. \eqref{gain}.
  (a) Gain as a function of $N$ (numerical and analytical solutions overlap).
  (b) Gain as a function of the input signal frequency, $\omega_d$, for a chain of $N = 10$ sites. 
  Analytical and numerical results are indistinguishable except at the divergence close to resonance at $\gamma_p = 2 t_d$.}
  \label{fig:Gain} 
\end{figure}

We assume now that the one-dimensional chain is driven by an input field at port $j=1$ with coherent and noisy components given by Eqs. (\ref{decompInput}, \ref{coherentInput}). Neglecting small non-amplified terms at $j=1$, we calculate the gain at the driving frequency $\omega_d$ and at photonic lattice sites with $j > 1$,
\begin{align}
G_j(\omega_d) &= \kappa^2 |Q_{j 1}(\omega_d)|^2 \nonumber \\  
                  &= G_1(\omega_d) e^{2 j /\xi(\omega_d)}
                  \nonumber \\
                  &= G_1(\omega_d)
                     \left(
                     \frac{(2 t_d)^2}{(\omega_d - \omega_0)^2 + (2 t_d - \gamma_p)^2}
                     \right)^j,
\end{align}
where
\begin{align}
G_1(\omega_d) &= \frac{\kappa^2}{\rho^2(\omega_d) s_0^2(\omega_d)} .
\label{gain}
\end{align}
Equation~\eqref{gain} shows that the output signal increases exponentially as a function of the position along the chain, as expected \tr{due to the effect} of directional amplification.
The gain, as expressed in the two last lines of Eq. \eqref{gain} is composed of two factors. 
The first one, $G_1(\omega_d)$, depends on $\omega_d$ with a typical width $2 t_d - \gamma_p$.
The second factor, $e^{2 j / \xi(\omega_d)}$, 
depends exponentially on $j$ and dominates the bandwidth gain, which for $j \gg 1$ can be approximated by
\begin{equation}
    \Delta \omega_d^{(j)} \approx \frac{2 t_d - \gamma_p}{\sqrt{j}} .
\end{equation}
We have checked that the analytical expression in Eq. \eqref{gain} agrees with numerical results obtained by calculating the matrix $Q_{jl}(\omega)$, as shown in Fig. \ref{fig:Gain}.

Finally, we can also calculate the gain in the total output signal, $S^{out}_T = \sum_j S^{out}_j = G_T(\omega_d) |\alpha|^2$, \tr{where the total gain reads,}
\begin{equation}
G_T(\omega_d) = \sum_j G_j(\omega_d)=  \kappa^2 \frac{e^{2 N / \xi(\omega_d)}}{\rho(\omega_d) s^2_0(\omega_d)} .
\label{gainT}
\end{equation}
In the last equation, we have neglected the corrections to the term $G_{j = 1}(\omega_d)$ from Eq.~\eqref{gain1} since the dominant contribution is from the exponentially amplified terms at $j > 1$. 

\subsection{Output noise}
We start by calculating the local spectral density of incoherent photons. We substitute \eqref{Q.edge} into \eqref{Nw}, and use the definition of $\xi(\omega)$ in \eqref{def.xi}
to get
\begin{align}
{}&n_j^{amp}(\omega)  = \kappa 
\frac{e^{2 j/\xi(\omega)}}{\rho(\omega) s_0^2(\omega)} 
\sum_{l,l'} 
\frac{1}{\rho(\omega)}
e^{-l/\xi(\omega)} e^{-{l'}/\xi(\omega)} 
\gamma^{(p)}_{l' l} 
\nonumber \\
 {}&
= \frac{4 \kappa t_d (1 + e^{-1/\xi(\omega)})}{\rho(\omega) s_0^2(\omega)} 
e^{2 j / \xi(\omega)}
\nonumber \\
 {}&
= \frac{4 \kappa t_d (1 + e^{-1/\xi(\omega)})}{\rho(\omega) s_0^2(\omega)} 
\left(
\frac{(2 t_d)^2}{(\omega\!-\!\omega_0)^2 + (\gamma_p\! -\! 2 t_d)^2} \right)^{j} .
\label{local.spectrum}
\end{align}
We have to integrate over frequency to obtain the form of the total output noise at each site. 
For this, we observe first that there are two factors in the expression for $n^{amp}_j(\omega)$ 
given in the last line of Eq. \eqref{local.spectrum}. They are both functions of $\omega$ with different bandwidths: $2 t_d  - \gamma_p$, and $(2 t_d - \gamma_p)/\sqrt{j}$, \tr{respectively}, as discussed below Eq.~\eqref{gain}. 
Thus, for large values $j \gg 1$, the following approximation is justified,
\begin{equation}
\int \frac{d\omega}{2 \pi} \ n_j^{amp}(\omega) \approx 
\frac{4 \kappa t_d \left( 1+e^{-1/\xi(\omega_0)}\right)}{\rho(\omega_0) s_0^2(\omega_0)} I(j),
\label{eq:topo_noise}
\end{equation}
where the integral reads,
\begin{align}
I(j) ={}& 
\int \frac{d\omega}{2 \pi} 
\left(
\frac{(2 t_d)^2}{(\omega-\omega_0)^2 + (2 t_d - \gamma_p)^2} \right)^{j} 
\nonumber \\
=
{}& \frac{2 t_d - \gamma_p}{2} 
\frac{1}{2^{2N-2}} \frac{(2 j - 2)!}{\left( (j - 1)! \right)^2}
\left( \frac{2 t_d}{2 t_d - \gamma_p} \right)^{2 j}
\nonumber \\
\stackrel{j \gg 1}{\approx}
{}& \frac{2 t_d - \gamma_p}{2 \sqrt{\pi}} \frac{1}{\sqrt{j-1}} 
\left( \frac{2 t_d}{2 t_d - \gamma_p} \right)^{2 j}.
\label{Ij}
\end{align}

\begin{figure}[t]
  \includegraphics[width=0.45\textwidth]{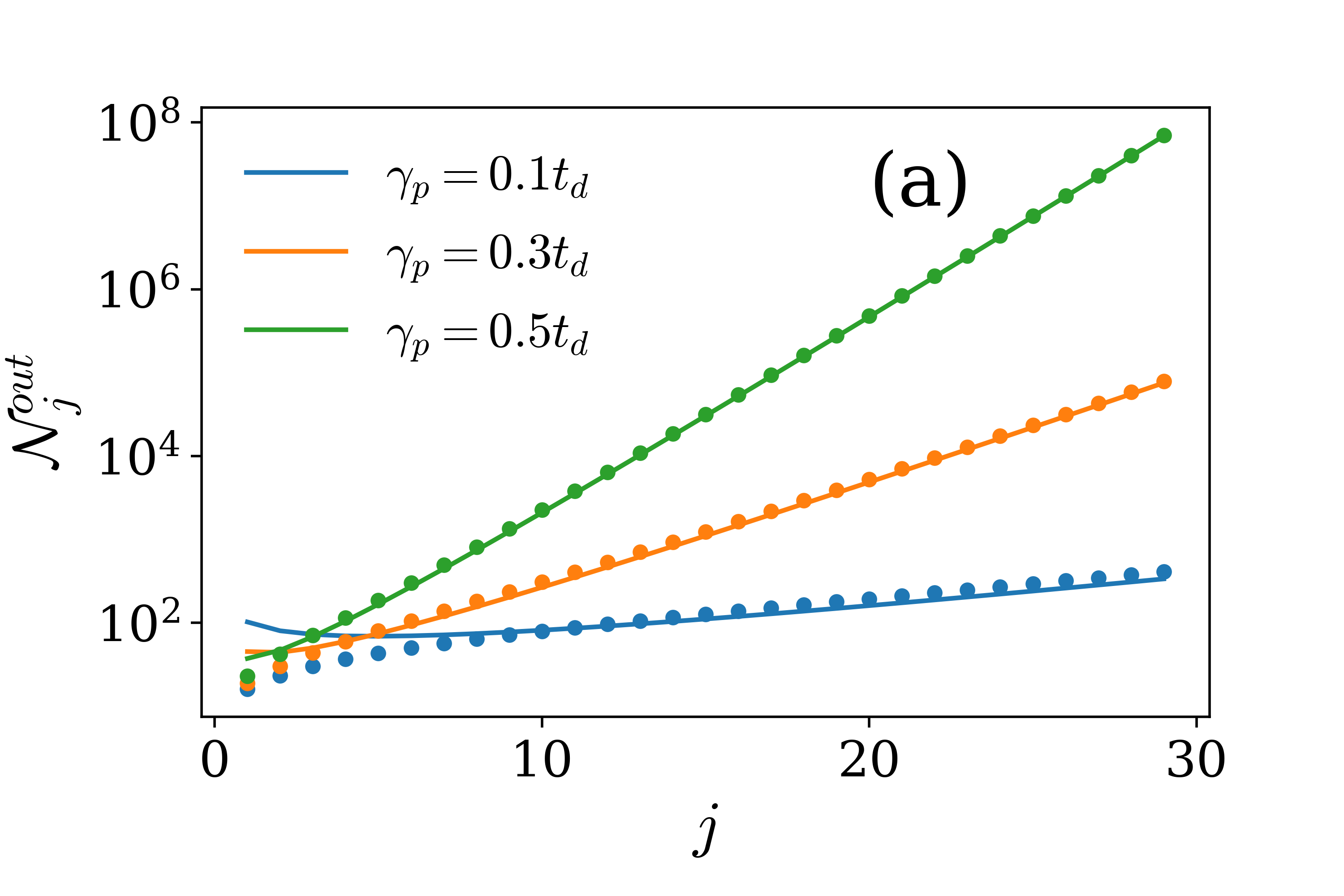}
  \includegraphics[width=0.45\textwidth]{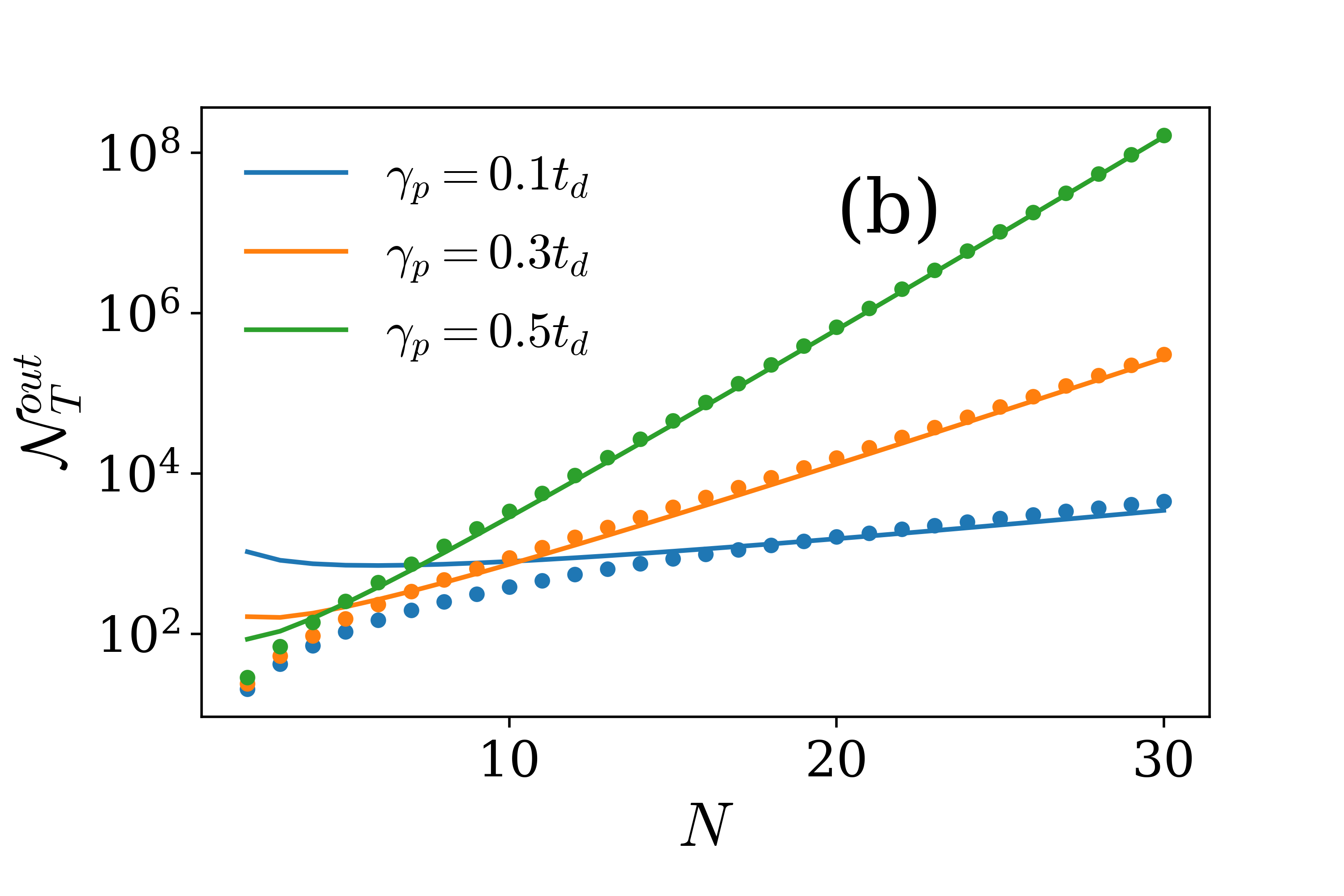}
  \caption{(a) Output noise as function of the lattice site of the photonic chain defined by Eq. \eqref{1D.couplings}, with $t_d = t_c$, $\phi = \pi/2$, $N = 30$ sites, and negligible input noise (${\cal N}^{in}_1 = 0$). Theory values predicted by the approximation in Eq. \eqref{photon_profile} are plotted as lines, whereas dots represent an exact calculation carried out by   numerically solving Eqs. (\ref{Nj}, \ref{Nw}). 
  (b) Same parameters as above, but we plot now the total output noise for photonic chains of different size.} 
  \label{Compare_N} 
\end{figure}

In the absence of any input field, or if we can neglect the noise component of the input field (${\cal N}_1^{in} = 0$), the previous expressions allow us to \tr{analytically} calculate the output noise, which can be finally written, at sites $j \gg 1$, as
\begin{eqnarray}
{\cal N}_j^{out} &=& \int \frac{d\omega}{2 \pi} \ n_j^{amp}(\omega) 
= {\cal N}^{amp}_0 \frac{1}{\sqrt{j-1}}  e^{2 j/\xi(\omega_0)} ,
\nonumber \\
{\cal N}^{amp}_0  &=& \frac{2 \kappa t_d (2 t_d - \gamma_p)(1 + e^{-1/\xi(\omega_0)})}{\sqrt{\pi} 
\rho(\omega_0) s_0^2(\omega_0)}.
\label{photon_profile}
\end{eqnarray}
Equation~\eqref{photon_profile} shows that incoherent photons decay exponentially up to a power-law correction. 
This implies that the edge-state localization length can be measured even in the absence of any coherent input, just measuring the distribution of the output noise along the chain.

Finally, we calculate the total output noise in the case \tr{of no input noise,} ${\cal N}_1^{in} = 0$, obtaining
\begin{align}
{\cal N}^{out}_{T} ={}& 
\sum_j {\cal N}^{out}_j 
\nonumber \\
\approx{}& \frac{2 \kappa t_d  (2 t_d - \gamma_p) (1+e^{-1/\xi(\omega_0)})}{\sqrt{\pi} s_0^2(\omega_0)} \frac{e^{2 N / \xi(\omega_0)}}{\sqrt{N-1}} . 
\label{NT}
\end{align}
In the last expression we have assumed that this expression can be evaluated in the limit $j \gg 1$, since those are the sites where noise is exponentially amplified.

A comparison between the analytical approximations in  Eqs. (\ref{photon_profile}, \ref{NT}) and exact numerical results is shown in Fig. \ref{Compare_N}. As expected, the approximation is more accurate as we increase $\gamma_p$ and the number of sites, and for values $\gamma_p/t_d \geq 0.5$ we observe almost perfect agreement.

\subsection{Added noise}
\begin{figure}[h]
  \includegraphics[width=0.50\textwidth]{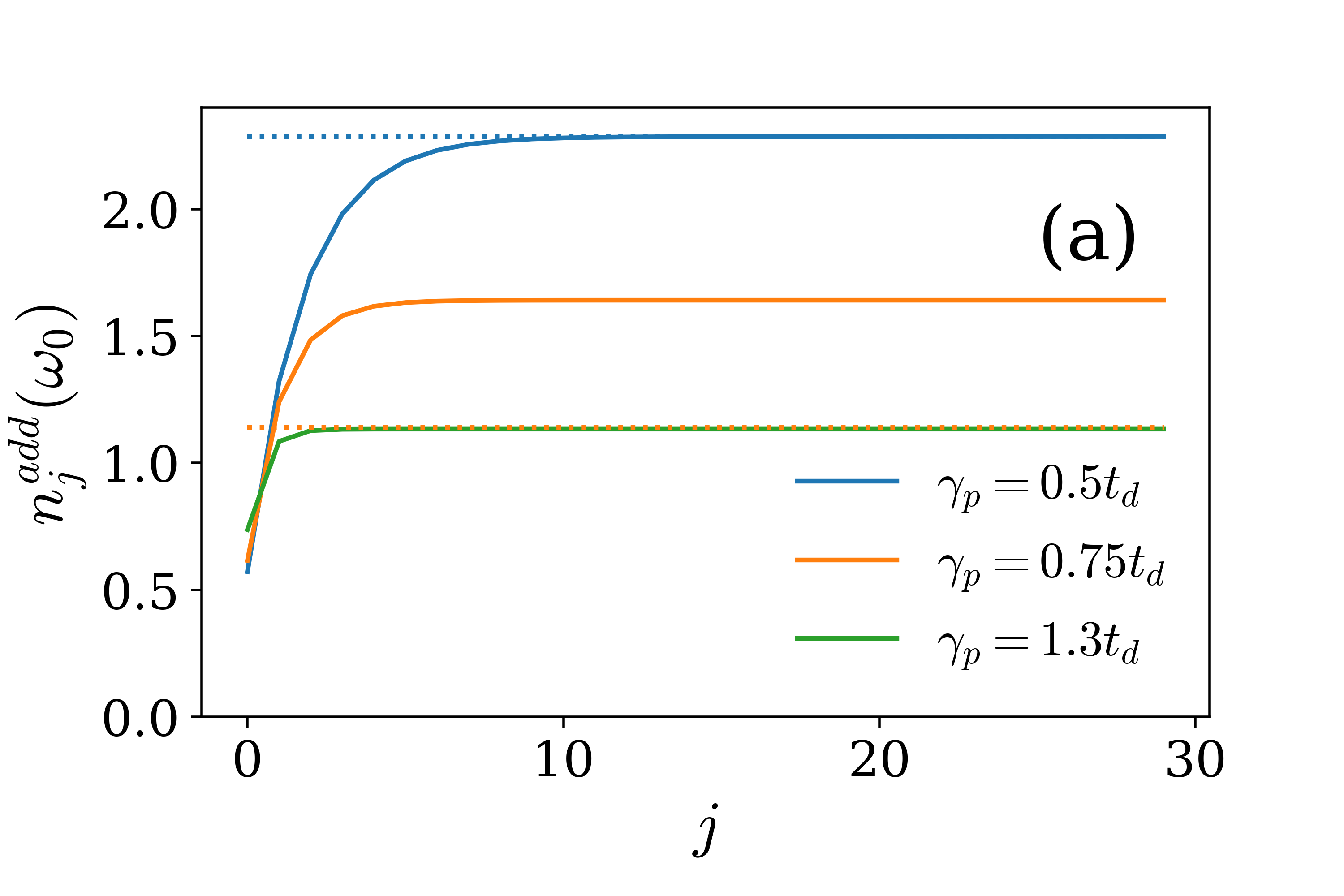}
  \includegraphics[width=0.50\textwidth]{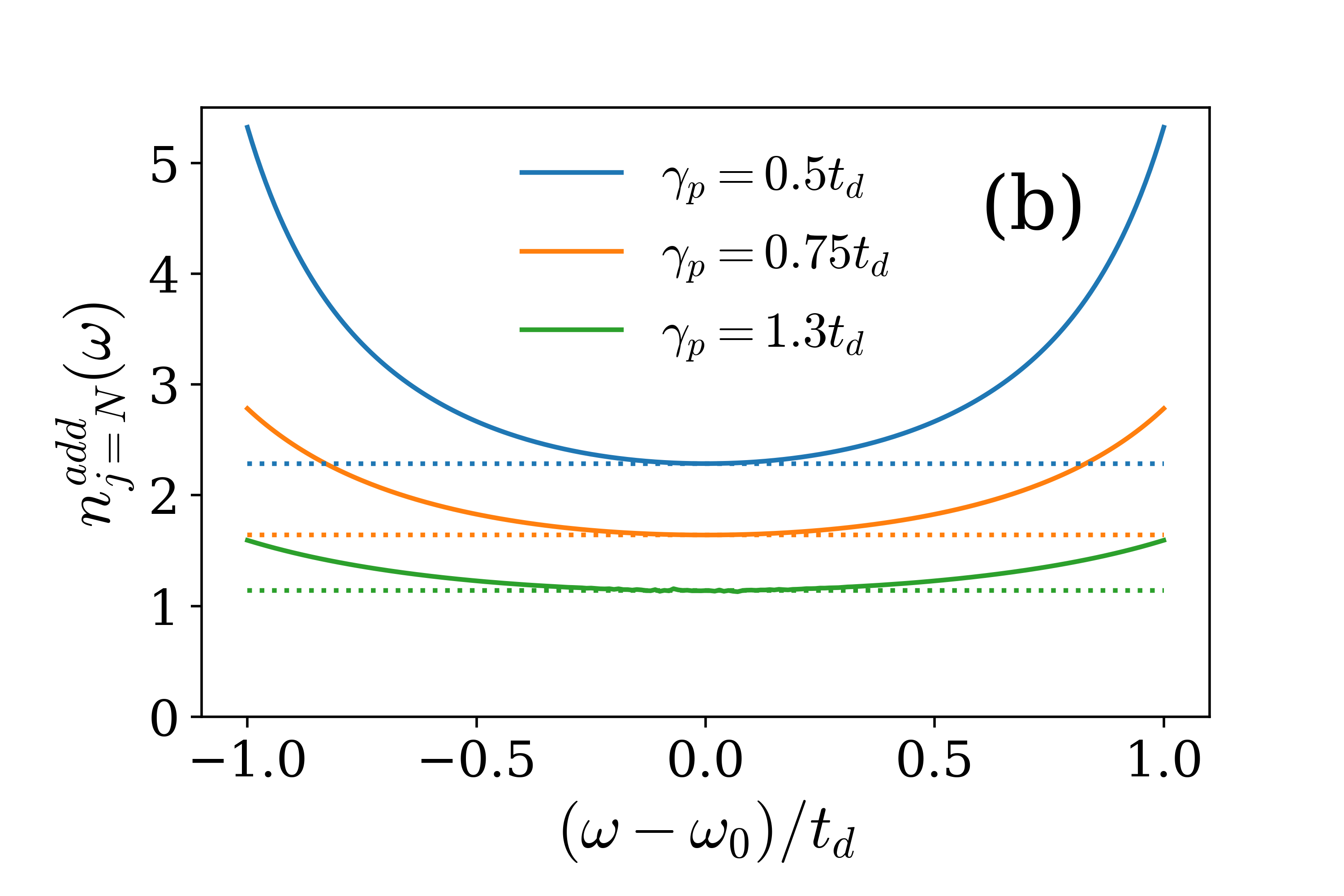}
  \caption{
  (a) Added noise at resonance ($\omega = \omega_0$) for a photonic chain with $t_d = t_c$, $N = 30$, and different values of the net photon pumping rate.  
  Continuous lines are the numerical evaluation of Eq. \eqref{addedNoise}, and dashed lines are the value given by 
  Eq.\eqref{added.noise.resonance} 
  (obtained under the condition that $j \gg 1 $). The plots show that the curves approach the value 
  $n_j^{add}(\omega_0) = 1$ as we increase $\gamma_p$. 
  (b) Same as above, but we plot now the added noise as a function of the frequency at the last site of the chain.} 
  \label{added_noise} 
\end{figure}

We can use the analytical results of the previous sections and calculate the added noise of the photonic lattice in the non-trivial topological phase. 
Evaluating Eq.~(\ref{addedNoise}), we get
\begin{align}
    n_j^{add}(\omega) &= \frac{4 t_d (1+e^{-1/\xi(\omega)})}{(1/\rho(\omega))\kappa}  
    \nonumber \\
    &= \frac{4 t_d}{\kappa} \frac{1}{1-e^{-1/\xi(\omega)}} \geq \frac{1}{2},
\label{added_noise_0}
\end{align}
\tr{valid for any large site $j\gg 1$.}
The inequality in Eq.~\eqref{added_noise_0} comes from the condition required for the driven-dissipative lattice to be in a non-trivial topological regime, that is $\gamma_p > 0$, leading to $4 t_d > \kappa/2$ from Eq.~(\ref{dissipative.couplings}). 
However, we find that in practice this quantum limit of added noise cannot be \tr{exactly} achieved in this model, because of the denominator $1-e^{-1/\xi(\omega)}$. To clarify this point, Let us consider the added noise at the resonant frequency $\omega = \omega_0$. Here, using expression 
Eq.~\eqref{def.xi}, we find
\begin{equation}
n_j^{add}(\omega_0) = \frac{8 t_d^2}{\kappa\tr{(2t_d-|\gamma_p-2t_d|)}},
\label{added.noise.resonance}
\end{equation}
\tr{where the topological non-trivial phase requires $0<\gamma_p<4t_d$ [see also Fig.~\ref{singularvaluegap} or Eq.~(\ref{cond.top})]. 
Notice that in the topologically trivial limits $\gamma_p \rightarrow 0$ or $\gamma_p \rightarrow 4t_d$, the added noise diverges
\begin{equation}
\lim_{\gamma_p \to 0, 4 t_d} n_j^{add}(\omega_0)  = +\infty.
\end{equation}
On the contrary, as we approach the center of the range $\gamma_p \rightarrow 2 t_d$  (or equivalently, $2 t_d \rightarrow \kappa/2$), the added noise reduces and reaches the limit,} 
\begin{equation}
\lim_{\gamma_p \to 2 t_d} n_j^{add}(\omega_0)  = 1.
\end{equation}
\tr{This is the minimum added noise that this simple SSH photonic lattice model for $j\gg 1$ can exhibit (see Fig.~\ref{added_noise} (a)) and corresponds to the limit where the edge-state localization length vanishes $\xi(\omega_0)\rightarrow 0$. For non-resonant frequencies $\omega\neq\omega_0$, the localization length increases $\xi(\omega)>0$ and the added noise is larger than at resonance, $n_j^{add}(\omega)>n_j^{add}(\omega_0)>1$, as shown in Fig.~\ref{added_noise} (b).}

Thus, the optimal situation in terms of the suppression of added noise is to be in regime of \tr{strongly localized edge-states $\gamma_p\rightarrow 2t_d$ and at resonant frequencies $\omega\rightarrow\omega_0$}.

\subsection{Noise-to-signal ratio}
We turn now to the the study of the noise-to-signal ratio. To simplify the analysis, we assume that the incoming signal is on-resonance with the cavity frequency, $\omega_d = \omega_0$, and also that the input field is at zero temperature  (${\cal N}_1^{in} = 0$). We compute the noise-to-signal ratio at every site $j \gg 1$ along the chain by using Eqs. (\ref{gain},\ref{photon_profile}), as well as the definition \eqref{def.xi}, \tr{obtaining}
\begin{figure}[h]
  \includegraphics[width=0.5\textwidth]{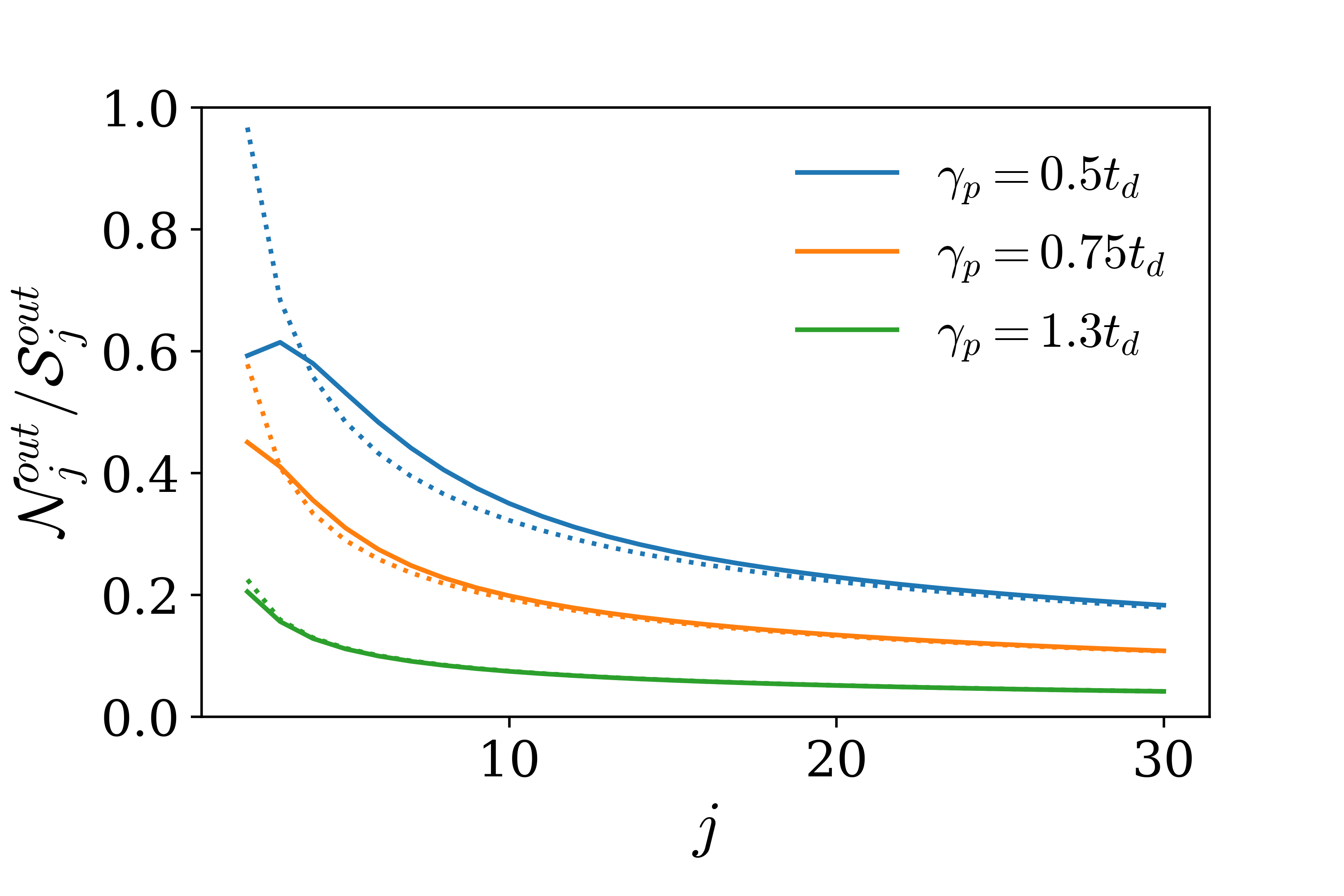}
  \caption{Noise-to-signal ratio, assuming ${\cal N}_1^{in} = 0$, as a function of the photonic lattice site, for a chain defined by \eqref{1D.couplings}, with $t_d = t_c$, $N = 30$, $\phi = \pi/2$, \tr{and $\omega_d=\omega_0$}. Continuous lines are an exact calculation, carried out by computing the output noise and signal by numerically calculating the matrix $Q(\omega)$ with Eq. \eqref{eq:Q}, and then using Eqs.~(\ref{Sjfull}, \ref{Sj}, \ref{Nw}, \ref{Nj}).} 
  \label{fig:noise-to-signal} 
\end{figure}
\begin{equation}
\frac{{\cal N}^{out}_j}{{\cal S}^{out}_{j}} =
\frac{4 t_d^2}{\kappa \gamma_p} \frac{(2 t_d - \gamma_p)}{|\alpha|^2} \frac{1}{\sqrt{\pi (j-1)}}.
\label{NtS1}
\end{equation}
We observe the remarkable feature that the noise-to-signal ratio decreases with the lattice site. \tr{Fig. ~\ref{fig:noise-to-signal} displays the exact numerical calculation of ${\cal N}^{out}_j/{\cal S}^{out}_{j}$ as well as its approximated analytical result (\ref{NtS1}) and confirms the validity of the scaling $\sim 1/\sqrt{j}$ for $j\gg 1$.} This dependence translates into the following expression for the total output noise-to-signal ratio, 
\begin{equation}
\frac{{\cal N}^{out}_T}{{\cal S}^{out}_{T}} =
\frac{4 t_d^2}{\kappa \gamma_p} \frac{(2 t_d - \gamma_p)}{|\alpha|^2} \frac{1}{\sqrt{\pi (N-1)}}.
\label{NtS2}
\end{equation}
We thus conclude that, in this model, increasing the size of the photonic chain not only leads to an exponential \tr{growth} of the gain, but it also leads to the suppression of the noise-to-signal ratio with a $\sim 1/\sqrt{N}$ scaling. 

\subsection{Stability of the topological amplification phases}

Let us now address the issue of stability of the dissipative phases of our one-dimensional example, \tr{given by Eqs.~\eqref{1D.couplings}}. Stability is a necessary condition for the model to be physically valid since -otherwise- fluctuations will lead to an increase of the photon number until, eventually, non-linearities become relevant.

\tr{As discussed in Sec.~\ref{section:stability}, a stable dissipative phase correspond to the case where all eigenvalues of the non-Hermitian matrix $H$ have negative real part. To check this in our model, we start analyzing the case of periodic boundary conditions, for which the eigenvalues of $H$ take the very simple form,}
\begin{equation}
\lambda(k) = \gamma_\pp - 2 t_\dd + 2 t_\dd \cos(k) -  i \omega_0 + 2 i t_\cc \cos(k + \phi).\label{eigenvaluesPeriodicBoundaries}
\end{equation}
\tr{Inspecting Eq.~(\ref{eigenvaluesPeriodicBoundaries})} we see that if the photonic chain is in a topologically non-trivial phase, \tr{given by conditions (\ref{conditions_topological}), then necessarily}  
$\Re (\lambda(k)) > 0$ for a at least some values of $k$, since otherwise the winding number associated to the vector $\tr{\vec{h}(k)=}(h_x(k),h_y(k) \tr{-} \omega)$ cannot take nonvanishing values. Thus, with the periodic boundary conditions, topological amplification is never stable. This is a very intuitive result, since in a periodic chain, any fluctuation is exponentially amplified \tr{without limit along the chain}.

The situation radically changes when we consider open boundary conditions. This is due to the well known {\it skin effect} present in non-Hermitian systems. 
This effect implies that eigenvalues of a non-Hermitian matrix can be very different for open or periodic boundary conditions, even in the large size limit. In the model \eqref{1D.couplings}, this can be easily checked, since we can diagonalize exactly the tridiagonal non-Hermitian matrix $H$ 
(see \cite{noschese13} for a derivation), obtaining

\begin{eqnarray}
\lambda_n &=& \gamma_\pp - 2 t_\dd 
\nonumber \\
&+&\!\! 2 \sqrt{(i t_\cc e^{i \phi} \! - \! t_\dd)(i t_\cc e^{-i \phi} \! -  \! t_\dd)} 
\cos\!\left( \! \frac{n \pi}{N \! + \! 1} \! \right)\!\!.
\end{eqnarray}
To simplify the discussion, we focus on the range of parameters that we have studied in detail in the previous subsections, namely, the case  $t_\cc = t_\dd$, $\phi = \pi/2$. Here, stable solutions exist if 
\begin{equation}
\gamma_p < 2 t_\dd. 
\label{eq:condition_stablity}
\end{equation}
This condition is clearly consistent with the existence of topological amplification phases as determined by Eq.~\eqref{cond.top}, and it is fulfilled by all examples studied in this work.

\subsection{Topological amplification under the effect of disorder}
Another important aspect of topological amplification phases is the role of disorder, which could be explored with the input-output scheme from this work. 
In this subsection we present numerical results that validate the intuition that non-trivial topological phases are robust against disorder.

\begin{figure}[t]
  \includegraphics[width=0.5\textwidth]{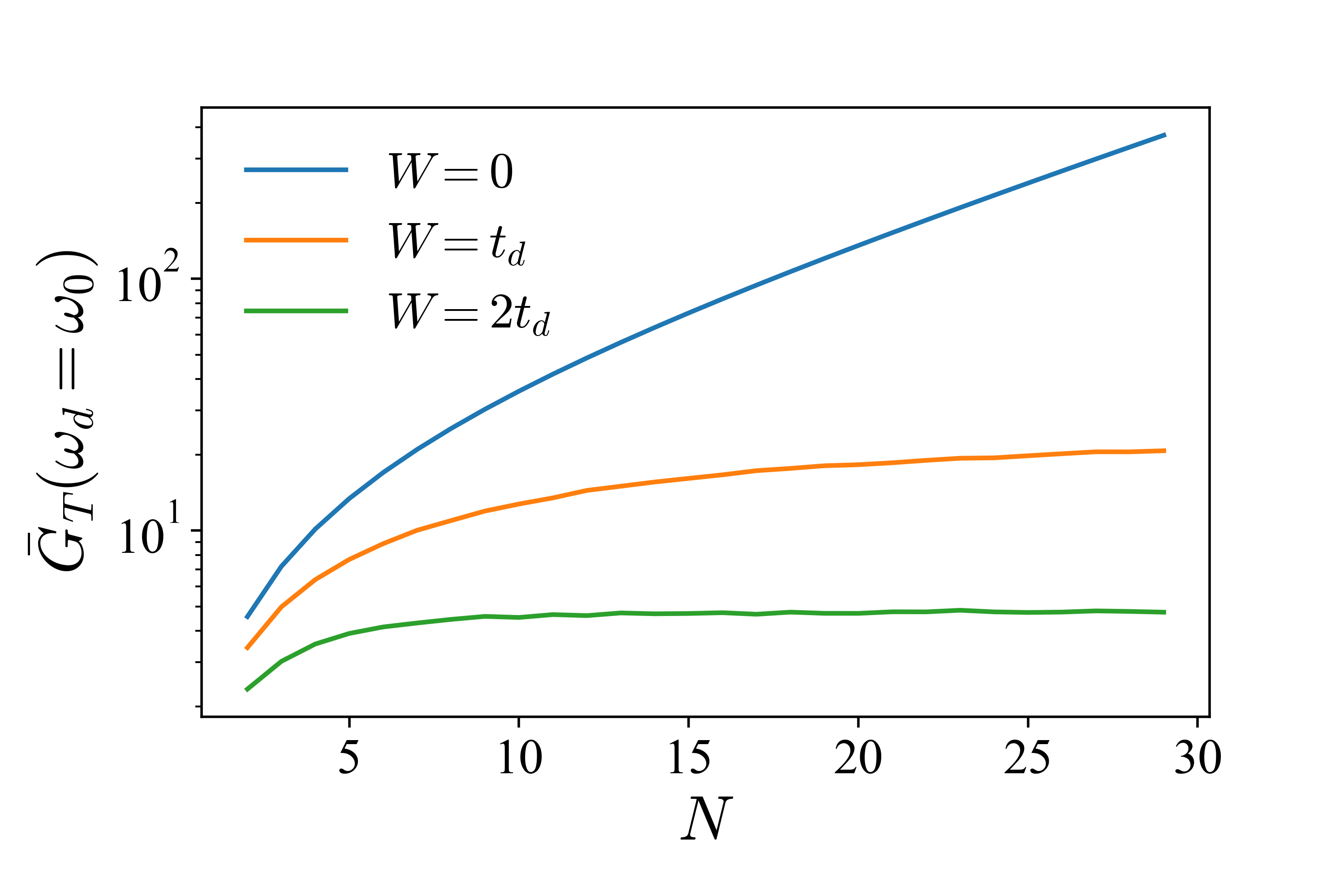}
  \caption{Average total gain at resonance as a function of the number of sites of the photonic chain \eqref{1D.couplings} as a function of $W$ (standard deviation the distribution of local mode frequencies), with values
  $t_d = t_c$, and $\gamma_p = 0.1$. Averages are taken over 5 10$^3$ instances of diagonal disorder.} 
  \label{gain_disorder} 
\end{figure}

We proceed by adding a diagonal disorder term to the Hermitian coupling matrix \tr{$H$ in Eq.~(\ref{def.H}). In particular, we consider local photonic modes with inhomogeneous resonance frequencies},
\begin{equation}
\omega_{j} = \omega_0 + \delta \omega_{j},
\end{equation}
where $\delta \omega_j$ are normal random variables with zero mean and standard deviation $W$.
This is well motivated physically, since many photonic lattices have a distribution of local mode energies due to imperfections in the fabrication process.
We use this model of diagonal disorder to calculate the average total gain, $\bar{G}_T = \sum_j \bar{G_j}$ as a function of the number of sites $N$, for different values of the disorder strength $W$, where $\bar{G_j}$ refers to the average of of $G_j$ over many instances of disorder.
\tr{As shown by our numerical results in Fig.~\ref{gain_disorder},} the exponential amplification effect survive until a finite value of the disorder strength is reached.

To investigate this dependence in a more quantitative way, we conjecture the following exponential dependence for the gain in the presence of disorder,
\begin{equation}
\bar{G}_T(\omega_0) \propto e^{m(W) N},
\label{def.mW}
\end{equation}
which is strongly supported by results like those presented in Fig. \ref{gain_disorder}.

\begin{figure}[t]
  \includegraphics[width=0.48\textwidth]{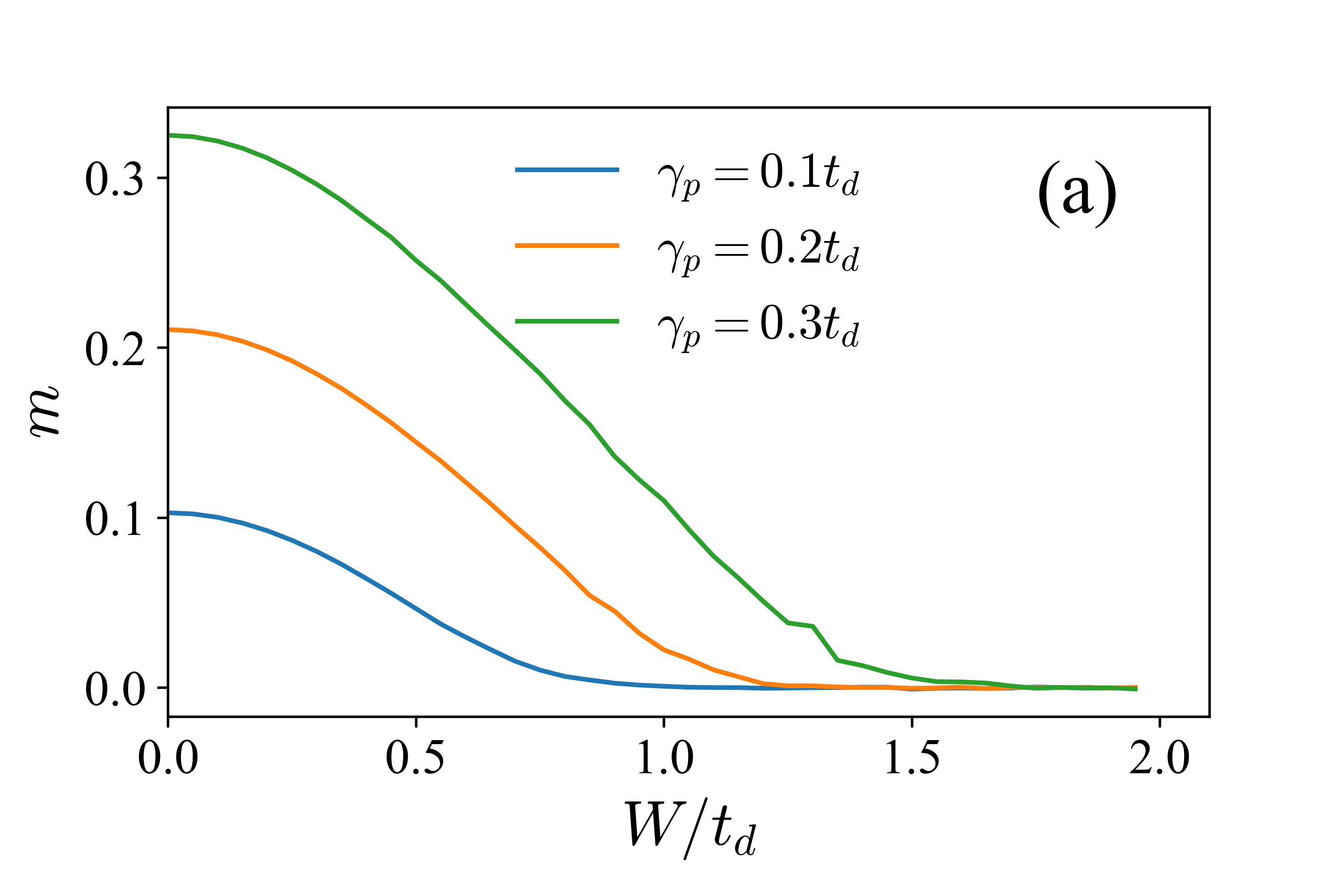}
  \includegraphics[width=0.48\textwidth]{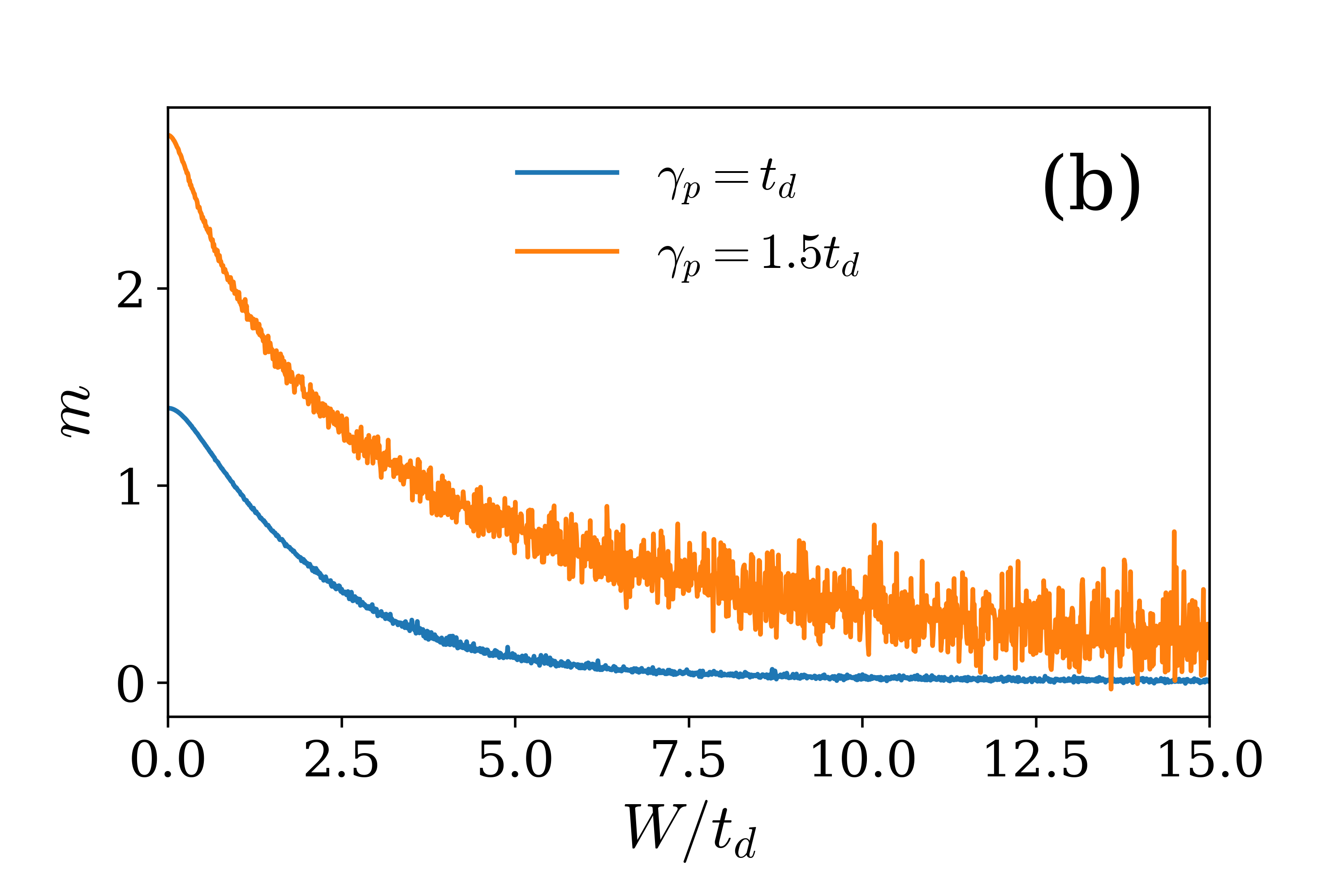}
  \caption{ 
  Exponent of the gain as defined by the ansatz in Eq. \eqref{def.mW}, calculated by fitting $\bar{G}_T$ curves like those shown in Fig. \ref{gain_disorder}. 
  Values are $t_d = t_c$, 
  $\phi = \pi/2$, and averages are taken over $5 \cdot 10^3$ disorder realizations.
  To obtain the disorder-dependent exponent, $m(W)$, we fit $\log(\bar{G}_T(\omega_0)) = m(W) N + a$, with $N = $ $40$, $50$, $\dots$, $120$.   (b) Same as before, but for higher values of the photon pumping rate. $\bar{G}_T$ curves are fitted now at the interval $N = 10, 11, \dots, 20$ and a number of 10$^4$ instances of disorder is used.} 
  \label{m_disorder} 
\end{figure}

Our numerical calculations show (see Fig.~\ref{m_disorder}) that \tr{there is a critical value of the disorder to reach in order to} make $m(W = 0)$ and \tr{thereby} break the topological amplification mechanism. 

\tr{Fig.~\eqref{m_disorder}(a) provides} indications of a disorder-induced phase transition at low values of $\gamma_p$ between a non-trivial topological phase and a non-amplifying phase, \tr{which} occurs as a function of disorder strength $W$. 
This transition seems to be smeared out as we increase the value of $\gamma_p$ and approach the value $2 t_d$, see Fig. \eqref{m_disorder}(b).
Further theoretical and numerical work is required to fully characterized this transition. \tr{Nevertheless,} since the \tr{critical} amount of disorder is \tr{on order} $W \approx t_d, t_c$, that is, \tr{comparable to} the photonic couplings, our results provide strong evidence that topological amplification is a robust mechanism \tr{against disorder} with a promising outlook for application in realistic devices, e.g. in superconducting circuit platforms.

\section{Conclusions and Outlook}
We have presented an input-output theory for \tr{topological many-body} photonic systems which relies on a connection between non-Hermitian coupling matrices and topological insulator Hamiltonians. Our results can be applied to any physical system that belongs to the \tr{broad} class of driven-dissipative bosonic lattices, including photonic and vibronic systems.
These ideas can be used to characterize the output signal \tr{and quantum noise} of 
non-reciprocal photonic lattices and directional amplifiers
 \cite{Devoret11,Devoret13,Devoret15,Metelmann15,Anderson16,Owens18,Lee16,Nunnenkamp18}.
The kind of driven-dissipative lattices considered here could be implemented in the quantum regime by breaking time-reversal symmetry by means of Floquet engineering in arrays of photonic or superconducting cavities \cite{Quijandria13prl,Peropadre13,navarrete14prl,Metelmann17,Quijandria18} or even trapped ion systems \cite{Bermudez11prl,Bermudez12,Kiefer19prl}. 

From a fundamental point of view, our work leads to a clear and unambiguous definition of topological phases and topological phase transitions in driven-dissipative bosonic systems.
\tr{The present theory can also be used to describe} topological non-trivial phases in the absence of any input signal, since we showed that the distribution of output noise along a photonic lattice is determined by the edge-states of the underlying topological insulator Hamiltonian. From a practical point of view, our work has a promising outlook in \tr{single-photon detection and near-quantum-limited amplification of quantum signals. In superconducting quantum circuits, for instance, directional amplification of microwave quantum signals could strongly improve the performance of traveling wave parametric amplifiers \cite{macklin_nearquantum-limited_2015,planat_photonic-crystal_2020,winkel_nondegenerate_2020,sivak_josephson_2020,renger_beyond_2020,malnou_three-wave_2020} and thereby increase the fidelity and signal-to-noise ratio of state-of-the-art qubit readout schemes \cite{gambetta_protocols_2007,jeffrey_fast_2014,walter_rapid_2017,dassonneville_fast_2020}.} Future promising research lines include the investigation of many-body effects \cite{fitzpatrick17} and non-linearities, for example in the case of cavity arrays in the lasing regime \cite{Fernandez18}. 

\section*{Acknowledgments}

We thank Alberto Cortijo for interesting discussions. 
Work funded by Spanish project PGC2018-094792-B-I00 (MCIU/AEI/FEDER, UE), CAM/FEDER project No. S2018/TCS-4342 (QUITEMAD-CM) and CSIC Quantum Technology Platform PT-001. T.R. further acknowledges support from the EU Horizon 2020 program under the Marie Sk\l{}odowska-Curie grant agreement No. 798397.

\appendix

	\section{Classification of topological amplification phases in terms of symmetries}
	\label{app:1}
	To classify the possible symmetry classes \tr{of the driven-dissipative lattice}, we conveniently rewrite Eqs.~(\ref{time},\ref{charge}). 
	In particular, we define Pauli operators transformed by unitary matrices $U_T$ and $U_C$ as, 
	\begin{align}
    \bar{\sigma}_\alpha {}&= U_T \sigma_\alpha U_T^\dagger, \nonumber \\
	\tilde{\sigma}_\alpha {}&= U_C \sigma_\alpha U_C^\dagger.
	\end{align}
	
    According to Eq.~\eqref{time}, time-reversal symmetry is fulfilled if there exist a unitary matrix $U_T$ such that,
	\begin{align}
	h_x(\vec{k}) \bar{\sigma}_x  - (h_y(\vec{k}) - \omega) \bar{\sigma}_y = h_x(-\vec{k})  \sigma_x +  (h_y(-\vec{k}) - \omega) \sigma_y.\label{T.symmetry}
	\end{align}
    \tr{Similarly,} invariance under charge conjugation can be expressed as
	\begin{align}
	h_x(\vec{k}) \tilde{\sigma}_x  - (h_y(\vec{k})-\omega) \tilde{\sigma}_y 
	= {}& 
	\nonumber \\
	- h_x(-\vec{k}) \sigma_x {}& -  (h_y(-\vec{k})-\omega) \sigma_y.
	\label{symmetry}
	\end{align}
Symmetry classes will be determined by
the range of possible unitary matrices
$U_T$ (since $U_C$ is subsequently determined by Eq.~(\ref{TC.condition})).
Note first that according to (\ref{symmetry}) $U_T$ has to generate a unitary transformation in the $x$-$y$ plane. Together with condition (\ref{restriction}), we find the following possibilities,
	\begin{eqnarray}
	U_T &=& e^{i \theta\sigma_z / 2}, \ U_C = e^{i (\theta+\pi/2) \sigma_z / 2},
	\nonumber \\
	U_T &=& \sigma_x, \ U_C = \sigma_y, \nonumber \\
	U_T &=& \sigma_y , \ U_C = \sigma_x .
	\end{eqnarray}
	Using those results and depending on the functions $h_x(\vec{k})$, $h_y(\vec{k})$, we can find the following possible symmetry classes  \cite{ryu10}:
	\begin{enumerate}[(i)]
		\item 
		$(h_x(\vec{k}))^2 +  (h_y(\vec{k}) - \omega)^2 \neq (h_x(-\vec{k}))^2 +  (h_y(-\vec{k})-\omega)^2$  
		$\to$ AIII class (no $T$, $C$ symmetry).
		\item 
		Vectors $(h_x(\vec{k}),-h_y(\vec{k})-\omega)$ and 
		$(h_x(-\vec{k}),h_y(-\vec{k})-\omega)$ are related by a rotation with angle $\theta$ on the $x$-$y$ plane $\to$ BDI class ($T^2 = C^2 = 1$) with $U_T = \exp(i \sigma_z \theta/2)$, $U_C = \exp(i \sigma_z (\theta+ \pi)/2)$.
		\item  $h_x(\vec{k}) = h_x(-\vec{k})$, $h_y(\vec{k}) - \omega= h_y(-\vec{k}) - \omega$ $\to$ CI class ($T^2 = 1$, $C^2 = -1$) with $U_T = \sigma_x$, $U_C = \sigma_y$.
		This is the case of real couplings matrices $\Gamma$, $G$.
		\item $h_x(\vec{k}) = - h_x(-\vec{k})$, $h_y(\vec{k})-\omega = - h_y(-\vec{k})-\omega$ $\to$ DIII class ($T^2 = -1$, $C^2 = 1$) with $U_T = \sigma_y$, $U_C = \sigma_x$.
	\end{enumerate}
    A remarkable aspect of this \tr{classification} is the fact that the particular symmetry class depends on the frequency $\omega$ of the incoming field. \tr{In addition, it} allows us to predict the existence or \tr{absence} of edge states. For example, in one dimension, non-trivial topological phases exist only in cases (i), (ii), (iv), which require the existence of complex photon tunneling terms or dissipative couplings.
	
	Let us see how this formalism applies to the particular one-dimensional lattice defined in Sec.~\ref{section:example}. We see from Eq.~\eqref{h.vector} \tr{that the conditions} 
	$h_x(k) = h_x(-k)$ and $h_y(k) = h_y(-k)$ \tr{are only fulfilled} if $\phi = 0, \pi$. Therefore, this dissipative system belongs to the AIII symmetry class unless $\phi = 0, \pi$, in which case it belongs to the topologically trivial CI class.
	
	\bibliographystyle{apsrev4-1}
	\bibliography{biblio}
	
\end{document}